\documentclass[aps,pra,twocolumn,groupedaddress, showpacs]{revtex4-1}
\usepackage[dvips]{graphicx}
\usepackage{bm}

\begin{document}

\title{Mapping the Berry Curvature from Semiclassical Dynamics in Optical Lattices}

\author{H. M. Price and N. R. Cooper}

\affiliation{TCM Group, Cavendish Laboratory, J. J. Thomson Ave., Cambridge CB3 0HE, United Kingdom} 

\bigskip

\bigskip

\begin{abstract}

We propose a general method by which experiments on ultracold gases can be used to determine the topological properties of the energy bands of optical lattices, as represented by the map of the Berry curvature across the Brillouin zone. The Berry curvature modifies the semiclassical dynamics and hence the trajectory of a wave packet undergoing Bloch oscillations. However, in two dimensions these trajectories may be complicated Lissajous-like figures, making it difficult to extract the effects of Berry curvature in general. We propose how this can be done using a ``time-reversal" protocol. This compares the velocity of a wave packet under positive and negative external force, and allows a clean measurement of the Berry curvature over the Brillouin zone. We discuss how this protocol may be implemented and explore the semiclassical dynamics for three specific systems: the asymmetric hexagonal lattice, and two ``optical flux" lattices in which the Chern number is nonzero. Finally, we discuss general experimental considerations for observing Berry curvature effects in ultracold gases.  
   
\end{abstract}


\maketitle

\section{Introduction}

One of the most interesting and surprising developments in the band
  theory of solids was the realization that the physical properties of any
  resulting energy band are not only encoded in its energy spectrum,
  $\varepsilon({\bf k})$, for all wavevectors ${\bf k}$ in the Brillouin zone (the
  ``band structure'' in the usual sense). In addition, there are important
  physical consequences related to the topology of the {\it eigenstates} that
  form the band \cite{hasankane,qizhang}.

  The importance of topological features of the energy eigenstates was first
  pointed out in the seminal work of Thouless {\it et al.} \cite{thouless} in
  the context of the integer quantum Hall effect. This work showed that, for
  two-dimensional (2D) lattices, the set of energy eigenstates that form an
  isolated band are characterized by an integer-valued Chern number, $C$, which can be nonzero when time-reversal symmetry is broken. The
  Chern number is a topological invariant of the band: its value cannot be
  changed by continuous evolution of the physical system without closing the
  gap to another band. Furthermore Ref.~\onlinecite{thouless} showed that the
  Chern number has direct physical consequences: a filled band has a quantized
  Hall conductance equal to $C$ times the quantum of conductance, corresponding to the existence of $C$ chiral edge modes.

  Soon after the work of Thouless {\it et al.}\cite{thouless}, the concept of
  the Berry phase was formulated \cite{berry}. The integer invariant of
  Ref.~\onlinecite{thouless} was quickly interpreted \cite{barrysimon} in terms of the
  integral of the Berry curvature in momentum space, $\Omega({\bf k})$, over the
  Brillouin zone. This Berry curvature, $\Omega({\bf k})$ (defined below), is a
  gauge-invariant property of the band structure which is predicted to have
  direct physical consequences when the band is partially
  filled \cite{niu2,haldanefs}. A full description of the properties of a nondegenerate
  energy band therefore requires a specification of both the spectrum,
  $\varepsilon({\bf k})$, and the Berry curvature, $\Omega({\bf k})$. For systems with additional global symmetries (time-reversal, 
particle-hole, or chiral) other forms of topological invariant can 
appear\cite{hasankane,qizhang}.

  These considerations have become of immediate importance in the field of
  ultracold gases in optical lattices. While the Berry curvature vanishes for the
  simplest forms of optical lattice, recent experiments have studied more
  complicated 2D optical lattices \cite{stamperkurnkagome,blochstaggered, esslinger}
  for which simple variants exist in which the (local) Berry curvature is nonzero.  Furthermore, there
  exist several theoretical proposals for forms of optical lattice in which
  the neutral atoms feel an effective magnetic field with high flux
  density \cite{JakschZoller,mueller,sorensen:086803,gerbier,dalibard,nigel,nigelnew}. These proposals are of great
  interest, as they offer the opportunity to study ultracold
  atoms in strong effective magnetic fields. These lattices break time-reversal symmetry in a
  way that leads to bands with nonzero Chern numbers \cite{JakschZoller,dalibard,nigel,
    nigelnew}.  A key motivation of this work is to propose how the
  bandstructures of these lattices might be characterized experimentally. We shall focus on the ``optical flux lattices'' proposed in
Refs.~\onlinecite{nigel,nigelnew}.

 A particle picks up a Berry phase when it travels adiabatically around a
  closed contour. Here, we are concerned with contours in two-dimensional momentum space, with
  ${\bf k}$ in the Brillouin zone. As usual, the Berry phase can be expressed
  as the integral of the Berry curvature, $\Omega({\bf k})$, over the surface
  bounded by the contour. If the integral is
  over the entire Brillouin zone, the periodicity in ${\bf k}$ space means
  that the closed contour is equivalent to a point, and the Berry phase must be quantized as $2 \pi C$,
  where $C$ is the integer-valued Chern number. This is the topological invariant
described above that underlies the remarkable quantization of the quantum
Hall effect \cite{thouless, kohmoto}. The Chern number has previously been measured through the conductance and chiral edge states of quantum Hall systems, and the total Berry phase associated with a Dirac point in graphene has been detected from transport \cite{novonobel,zhangnature} and ARPES \cite{liu} measurements. To our
knowledge, the local Berry curvature has not been measured
directly, although its influence has been detected on a number of physical
quantities \cite{di}. It also plays an important role in the anomalous quantum
Hall effect \cite{jungwirth,onoda, haldanefs} and in semiclassical dynamics \cite{niu2,
  chang,sundaram}.

Ultracold gases present an excellent opportunity to study the Chern
number and the topology of bands directly. Optical lattices can impose
periodic potentials, and ultracold gas experiments are highly controllable,
tunable, and clean. Recent theoretical papers have shown how to detect the
Chern number for certain tight-binding models in time-of-flight measurements
\cite{spielman, pachos}. In this paper, we propose a much more general method,
based on semiclassical dynamics, that can be used to directly map out the
Berry curvature over the Brillouin zone for any form of underlying lattice.

In the absence of Berry curvature, a wave packet subjected to a constant
external force will execute Bloch oscillations. These have not been observed
for bulk crystalline electrons due to electronic scattering off lattice
defects, but Bloch oscillations have been seen in other physical systems
including semiconductor superlattices \cite{feldman} and ultracold gases
\cite{dahan,anderson}.

In the presence of Berry curvature, the wave packet dynamics will be strongly
modified. In principle, these effects may be measured directly from the
real-space trajectory of such a wave packet, as has been previously proposed
\cite{dudarev, diener}. However, in two dimensions, Bloch oscillations are
complicated \cite{korsch, witthaut, mossmann}, and it can be difficult to
disentangle the effects of Berry curvature from the usual effects arising from the group velocity. We propose how this may be overcome using a ``time-reversal" protocol. This will allow experiments to map the Berry curvature over the entire Brillouin zone. 

The structure of the paper is the following. First,  we introduce the Berry curvature and describe its effects on the semiclassical dynamics. We then discuss the complications of 2D Bloch oscillations and outline a protocol for how the Berry curvature may be mapped experimentally. We illustrate this discussion with numerical results for the Berry curvature and semiclassical dynamics for three interesting specific models: the asymmetric honeycomb lattice and two optical flux lattices  \cite{nigel, nigelnew}. Finally we discuss general experimental considerations and time-of-flight experiments.

\section{The Berry Curvature and Semiclassical Dynamics}

From Bloch's theorem, the eigenfunctions of a periodic potential can be expressed as $\psi_{n,{\bf k}}({\bf r}) =  u_{n,{\bf k}}({\bf r})e^{i {\bf k} \cdot {\bf r}}$, where the Bloch function, $u_{n,{\bf k}} ({\bf r})$, has the periodicity of the underlying lattice and $n$ is the band index. We confine our discussion to the 2D $xy$ plane, where the band structure is characterized by the energy, $\varepsilon_n({\bf k})$, and the scalar Berry curvature, $\Omega_n({\bf k})$. All the concepts can be extended to three dimensions, where the Berry curvature must be treated vectorially. 

The Berry phase, $\gamma_n$, for adiabatic transport in ${\bf k}$ space around a closed curve $\mathcal{C}$ bounding a region $\mathcal{S}$ is
\begin{eqnarray} 
\gamma_n &=&  \oint_\mathcal{C} d {\bf k} \cdot {\bf A}_n ({\bf k})=\int_\mathcal{S} d{\bf S} \cdot \hat{{\bf z}} \Omega_n({\bf k}) , \label{eq:berry}\\
{\bf A}_n ({\bf k}) & \equiv & i \langle u_{n,{\bf k}}| \frac{\partial}{\partial {\bf k}} |u_{n,{\bf k}}\rangle , \\
\Omega_n({\bf k})& \equiv & {\bf \nabla_k} \times {\bf A}_n ({\bf k}) \cdot\hat{{\bf z}} , \label{eq:omega}
\end{eqnarray}  
where ${\bf A}_n ({\bf k})$ is the Berry connection, a gauge-dependent vector
potential, and $\Omega_n({\bf k})$ is the Berry curvature. The Berry curvature is a gauge-invariant,
local property of the band structure. It vanishes when both time-reversal and
inversion symmetries are present. The Berry phase is geometrical and similar
to the Aharonov-Bohm phase, with the Berry curvature playing the role of a
magnetic field. When integrated over the whole Brillouin zone, the resultant Berry phase is quantized and equal to $2 \pi$ times the Chern number. The Chern number is thus a global topological property of the band, and underlies phenomena such as the integer quantum Hall effect \cite{thouless}. However, the local Berry curvature also has important physical consequences, for instance, on the semiclassical motion of a wave packet \cite{niu2, chang,sundaram, di}.        

To describe the semiclassical dynamics, we consider a gas of noninteracting fermions or bosons that is prepared in a wave packet with a center of mass at position ${\bf r}_c$ and momentum ${\bf k}_c$ \cite{ashcroft, dahan}. For atoms initially prepared in the bottom of the lowest band, the temperature contributes to the initial momentum spread of the atoms. We therefore assume that the temperature is less than the bandwidth so that the wave packet does not cover the whole Brillouin zone. We then consider a constant external force ${\bf F}$, which in a solid state system would usually be due to an electric field. However, ultracold gases are neutral, and this force instead may come from linearly accelerating the lattice \cite{dahan, bharucha} or from gravity \cite{anderson, roati, fattori, ferrari}. It is assumed that the force is sufficiently small that the motion is adiabatic such that Landau-Zener tunneling\cite{zener} is negligible and the wave packet remains in a single band; henceforward we drop the band subscript $n$. The semiclassical equations of motion are then \cite{chang}
\begin{eqnarray}   
\dot{{\bf r}}_c = \frac{1}{\hbar} \frac{\partial \varepsilon ({\bf k}_c)} {\partial {\bf k}_c} -  (\dot{{\bf k}}_c \times \hat{{\bf z}}) \Omega ({\bf k}_c), \label{eq:rc}\\
 \hbar \dot{{\bf k}}_c = {\bf F} . \label{eq:motion}
\end{eqnarray}  
We shall refer to the first term in (\ref{eq:rc}) as the group velocity and to the second as the Berry velocity. 
We note that these semiclassical equations are further modified if there is an external ``magnetic" field \cite{chang} in addition to the external force ${\bf F}$. We shall not discuss this further here, assuming that any magnetic field has the periodicity of the lattice and is incorporated into the bandstructure through the (magnetic) Bloch states $u_{{\bf k}}({\bf r})$ (see Sec. \ref{sec:ofl}). The effect of a magnetic field on Bloch oscillations has also been analyzed directly from the tight-binding Hamiltonian in Refs. \onlinecite{mantica, kolocyclo}.  

To theoretically simulate the semiclassical dynamics, we must be able to calculate the Berry curvature. In general, a simple analytic expression is not possible and the Berry curvature is calculated numerically. This requires a discretized version of (\ref{eq:omega}), as eigenfunctions are found computationally over a grid in ${\bf k}$ space. There is an inherent phase ambiguity in the Bloch states, and so a gauge must be chosen to calculate the Berry connection. The Berry curvature is gauge invariant, and can be found on this grid by the method of Fukui {\it et al.} \cite{fukui}, which applies a geometrical formulation of topological charges in lattice gauge theory, where the Berry curvature is calculated from the winding of $U(1)$ link variables around each plaquette in the Brillouin zone. We have used this method throughout this work for the numerical calculations.

It is of interest to note that effects of Berry curvature also arise in the semiclassical dynamics of a wave packet in a time-dependent one-dimensional optical lattice \cite{di, kitagawa, pettini}. The Berry curvature is then defined over a 2D parameter space made up of the one-dimensional quasimomentum and time. The Bloch oscillations of a wave packet in such a potential were theoretically studied in Ref.~\onlinecite{pettini}. 

\section{2D Bloch Oscillations} \label{sec:lissa}

The pioneering experiments on Bloch oscillations in ultracold gases were (quasi-)one dimensional \cite{dahan, anderson}, and only recently has the extension to 2D been investigated \cite{esslinger}. In 2D, Bloch oscillations have various interesting features in their own right, even before the Berry curvature is considered. 

One important consequence of dimensionality is that the real-space Bloch oscillations in 2D become Lissajous-like \cite{witthaut, korsch, mossmann}. For separable potentials, 1D Bloch oscillations along the $x$ and $y$ axes are simply superposed. For an arbitrary force ${\bf F}=(F_x,F_y)$, the wave packet's motion is periodic along $k_i$ with periods  $\tau_{Bi}=  h / |F_i| a$ (where $i$ runs over $x$,$y$). The ensuing motion depends on the ratio $F_x : F_y$. For nonseparable potentials, studies show that similar behavior can be expected when the force applied is weak and Landau-Zener tunneling is negligible \cite{witthaut, korsch}.

\begin{figure} [htdp]
\centering
\resizebox{0.20\textwidth}{!}{\includegraphics*{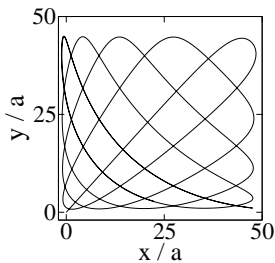}}  
\caption{An example of a Lissajous-like figure for the square optical flux lattice (Sec. \ref{sec:sq}). The ratio $F_x:F_y$ is 9 : 10, and a low force of $|{\bf F}|=0.05 F_R$ is used to minimize the effects of Berry curvature over one oscillation ($F_R = h^2/2m \lambda^3$ where $\lambda$ is the optical wavelength and $m$ is the mass of the atomic species; see Sec. \ref{sec:system}). The Lissajous-like figure is approximately bounded by the Bloch oscillation lengths, and so it obscures the effects of Berry curvature within this box.} \label{fig:lissa}
\end{figure}      

The real-space Lissajous-like figures can be complicated two-dimensional oscillations, bounded by $x_{Bi} \propto v_{Bi}\tau_{Bi}$ (Fig. \ref{fig:lissa}). For the Berry curvature to change this trajectory significantly, it would be necessary to wait until the wave packet drifts outside of the bounding box. As a result, experiments would measure only the net Berry curvature encountered along a path. Information would be lost about how the Berry curvature is distributed in momentum space and notably whether its sign changes.   

Furthermore, in 2D there can be an additional drift in the wave packet's position, independent of the Berry curvature, if the wave packet does not start at high symmetry points such as the zone center ${\bf k}_0=(0,0)$ \cite{mossmann, zhang}. Thus, merely observing a transverse drift is not, by itself, conclusive evidence of nonzero Berry curvature.

\section{A ``Time-Reversal" Protocol to Extract The Berry Curvature}

Berry curvature effects can be isolated by considering the dynamics under a reversal of ``time." In doing so, it is important to be able to measure the velocity of the wave packet. We shall discuss in Sec. \ref{sec:ex} how this may be done in experiments: for instance, through tracking the position of the wave packet in real space or through the momentum distribution, as in the seminal paper of Ben Dahan {\it et al.} \cite{dahan}.

We consider first measuring the velocity for a given force, ${\bf F}$, at a particular point, ${\bf k}$, in the Brillouin zone. This can be achieved in an experiment in which the wave packet has evolved according to ${\bf k}(t)={\bf k}(0)+ {\bf F} t /\hbar$. This velocity is uniquely defined (within the single band approximation) at each point ${\bf k}$ along the trajectory and we denote it as ${\bf v}_{{\bf k}} (+{\bf F})$. We now consider measuring the velocity in an experiment in which the wave packet passes through the same point ${\bf k}$, but with opposite direction of the force, $-{\bf F}$, which we denote ${\bf v}_{{\bf k}} (-{\bf F})$.  This can be achieved, for example, by evolving the wave packet along the line ${\bf k}(0)+ {\bf F} t /\hbar$ for a time $T$ that moves it past the point of interest (e.g., to the Brillouin zone boundary), and then retracing this path using the reversed force $-{\bf F}$.
 From (\ref{eq:rc}), we can see that the Berry velocity changes sign, while the group velocity remains invariant. The two effects can then be separated:
\begin{eqnarray}   
{\bf v}_{\bf_k}(+{\bf F})-{\bf v}_{\bf_k}(-{\bf F})= -\frac{2}{\hbar}({\bf F} \times \hat{{\bf z}}) \Omega({\bf k}),\\
{\bf v}_{\bf_k}(+{\bf F})+{\bf v}_{\bf_k}(-{\bf F})= \frac{2}{\hbar}  \frac{\partial \varepsilon({\bf k})} {\partial {\bf k}} .
\end{eqnarray}  
This transformation is equivalent to a time-reversal operation, and it cleanly removes the effects of the complex Lissajous-like figures in 2D. 

\begin{figure} [htdp]
\centering
$
\begin{array}{cc}
(a)\resizebox{0.16\textwidth}{!}{\includegraphics*{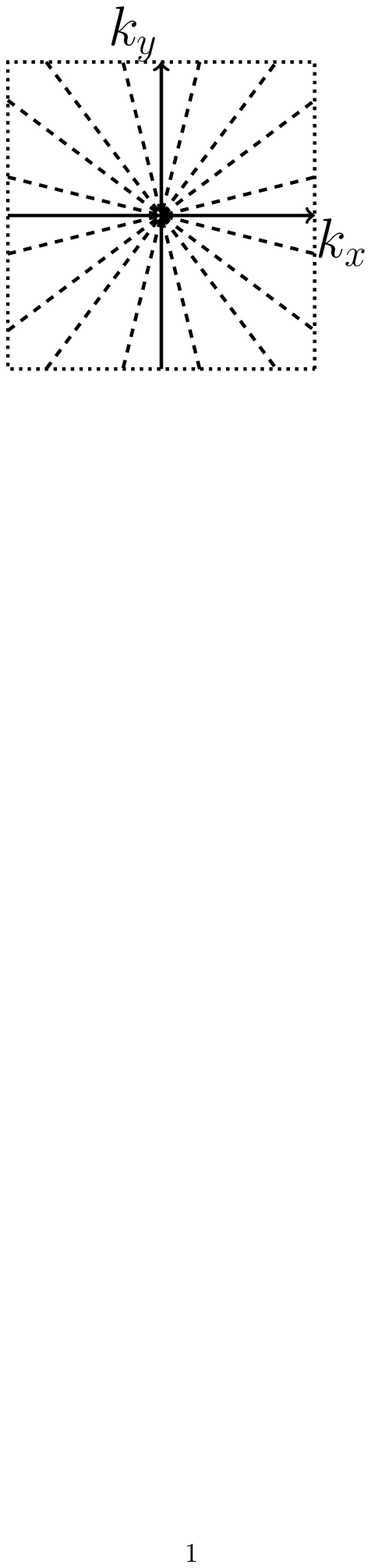}} &
(b)\resizebox{0.16\textwidth}{!}{\includegraphics*{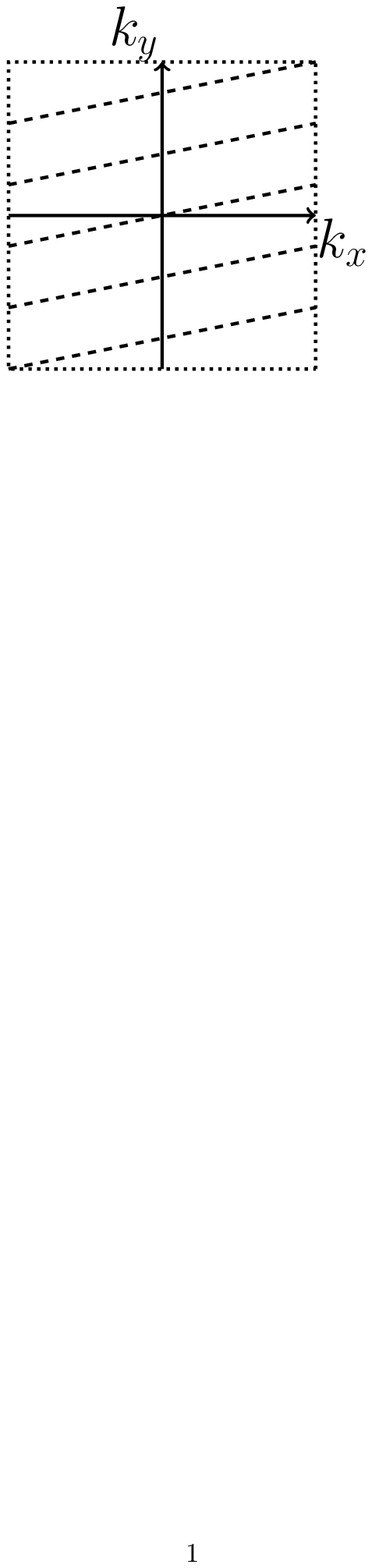}}  
\end{array}$
\caption{Two methods for mapping the Brillouin zone. (a) Rotating the force with respect to the lattice between experiments allows each wave packet to travel a different path. (b) With a large ratio $F_x : F_y$, a single wave packet successively travels many paths within the Brillouin zone.} \label{fig:bzmap}
\end{figure}      

The Berry curvature can now be found at each point along the wave packet's path. By varying the path across the whole Brillouin zone, the Berry curvature is mapped out and the Chern number is directly measured. The path of the wave packet may be chosen in various ways, two of which are illustrated in Fig. \ref{fig:bzmap}. First, the alignment of the force with the lattice may be rotated so that different trajectories are successively explored. In such a scheme, the measurement time can be short, corresponding to the time taken for the wave packet to travel once across the Brillouin zone. However, it would be important to align the force precisely each time. An alternative scheme would be to make the ratio $F_x:F_y$ large. The wave packet will cover a large area of the Brillouin zone during a single Bloch oscillation. The force needs to be aligned only twice (for $+{\bf F}$ and $-{\bf F}$) but longer measurement times would be required. A combination of these methods may be most suitable.

\subsection{Relation to the Chern number} \label{sec:relchern}

By this approach measurements of the velocity of the wave packet can be used to extract the Berry curvature. Measurements of the net drift of the wave packet in position space can be used to measure the Chern number of the band. To illustrate the idea, it is convenient to consider a Brillouin Zone (BZ) of rectangular symmetry, and the set of paths ${\bf k} = (0,k_y) \to (K_x,k_y)$ that are traced out by a force $+F$ in the $x$ direction, for different values of the initial wave vector $k_y$.  Here $K_x$ denotes the size of the reciprocal lattice vector in the $x$ direction, so the path traverses the full width of the Brillouin zone once. The net drift of the wave packet in the $y$ direction, $\Delta y = \int v_y dt$, is
\begin{equation}
\Delta y(k_y, +F) = \int_0^{K_x} \frac{1}{ F}\frac{\partial \varepsilon}{\partial k_y} dk_x +
\int_0^{K_x} \Omega(k_x,k_y) dk_x .
\end{equation}
Note that, for general $k_y$, there is a transverse displacement not only from the Berry curvature but also from the group velocity \cite{mossmann, zhang}. Reversing the force, such that the set of paths run in the opposite direction, over ${\bf k} = (0,k_y) \to (-K_x,k_y)$, the displacement becomes
\begin{equation}
\Delta y(k_y, -F)   =  \int_0^{K_x} \frac{1}{ F}\frac{\partial \varepsilon}{\partial k_y} dk_x -
\int_0^{K_x} \Omega(k_x,k_y) dk_x .
\end{equation}
Thus, the contribution from the group velocity stays the same, but the contribution from the Berry curvature changes sign. Averaging the difference,
\begin{equation}
\frac{\Delta y(k_y, +F) -\Delta y(k_y, +F) }{2}= \int_0^{K_x} \Omega(k_x,k_y) dk_x 
\end{equation}
extracts the part that depends on the Berry curvature. It is interesting to note that this contribution is independent of the magnitude of the force. The size of the transverse displacement is just set by the length scale of the underlying lattice (the lattice constant), and a numerical factor that involves the average Berry curvature along the trajectory.  Furthermore, the total Chern number can be found by summing over the set of trajectories with different values of $k_y$, which can be used to represent an evaluation of the integral:
\begin{equation}
 C =\frac{1}{2\pi} \int_0^{K_y} d k_y \int_0^{K_x} dk_x \Omega(k_x,k_y) 
 \end{equation}
at discrete points in $k_y$. (Clearly this approach can be readily adapted to a lattice of any symmetry, provided the set of paths spans the entire Brillouin zone once.)

Note that, if instead of a wave packet, the band is filled (e.g., by noninteracting fermions) the Chern number may be measured from the net current density when a force, ${\bf F}$, is applied: 
\begin{equation}
{\bf J} = \frac{1}{2 \pi h}  \int_0^{K_y} d k_y \int_0^{K_x} dk_x \Omega(k_x,k_y) ({\bf F} \times \hat{{\bf z}})=\frac{C}{h}({\bf F} \times \hat{{\bf z}}) .
\end{equation}
For a trapped gas, this result can be applied locally, with ${\bf F}$ set by the local potential gradient to give equilibrium currents. 

\section{Example Systems} \label{sec:system}

In this section, we illustrate our proposed method for measuring the Berry curvature for three example systems that are of experimental interest: the asymmetric hexagonal lattice, and two optical flux lattices \cite{nigel,nigelnew} for which the Chern number is nonzero. 

In optical lattices, the natural energy scale is set by the recoil energy, $E_R=h^2/2m \lambda^2$, where $\lambda$ is the optical wavelength. Similarly, we can define a recoil velocity, $v_R=h/m\lambda$, and a unit of force, $F_R=h^2/2m \lambda^3$. Hence forward, we express all quantities in these units.

The magnitude of the external force significantly affects the dynamics, as discussed further in Sec. \ref{sec:ex}. In previous experiments, the force has been introduced by linearly accelerating the lattice \cite{dahan, bharucha}, where the magnitude can be varied, or by gravity \cite{anderson, roati, fattori, ferrari}. In our units, $|m {\bf g}|=0.7 F_R$ for $^{174}$Yb and for $\lambda=\lambda_0\approx578$ nm, the resonance wavelength coupling the ground and excited state in $^{174}$Yb \cite{gerbier}. This choice of parameters is especially relevant to the optical flux lattices discussed below. We therefore primarily focus on the representative case $|{\bf F}| = 1 F_R$.   

For the evolution of the wave packet to be adiabatic, the rate of Landau-Zener tunneling to the next lowest band must be small. The probability of a Landau-Zener transition where the bands almost touch is given by \cite{zener}
\begin{equation}
p = e^{-a_c/a} \label{eq:zener},
\end{equation}
where $a$ is the acceleration of the atoms moving under the external force, $a_c=(\delta \varepsilon)^2 \lambda / 8 \hbar^2$, and $\delta \varepsilon$ is the band gap. This can therefore be neglected when the force is small or the band gap is large. 

\subsection{The asymmetric hexagonal lattice} \label{sec:hex}

The tight-binding hexagonal lattice has long been studied in condensed
matter physics as a simple model for graphene \cite{semenoff}. Thanks
to recent advances, optical lattices of hexagonal symmetry (or
  closely related forms) can be imposed on ultracold gases and
  phenomena associated with the interesting band topology can be
directly studied \cite{hexagonal,esslinger, montambaux}.

In the presence of both inversion and time-reversal symmetry, the bands touch at two Dirac points in the corners of the hexagonal Brillouin zone. If either of these symmetries is broken, band gaps open and Berry curvature appears at these points, as in the Haldane model where time-reversal symmetry is broken \cite{haldane}. The Chern number has also been observed experimentally for time-reversal symmetry breaking in graphene \cite{zhangnature, novonobel}.  

The asymmetric hexagonal lattice instead breaks inversion symmetry by introducing an onsite energy difference between the two lattice sites, $A$ and $B$. The opening of band gaps with asymmetry has already been studied experimentally in graphene \cite{zhou2}, but Berry curvature effects have not been observed directly. Theoretically, the Berry curvature can lead to a quantum valley Hall effect, which may be useful for valley-based electronic applications \cite{semenoff, xiao, di}. It would therefore be of great interest to study this system in ultracold gases.      

\begin{figure} [htdp]
\centering
\resizebox{0.35\textwidth}{!}{\includegraphics*{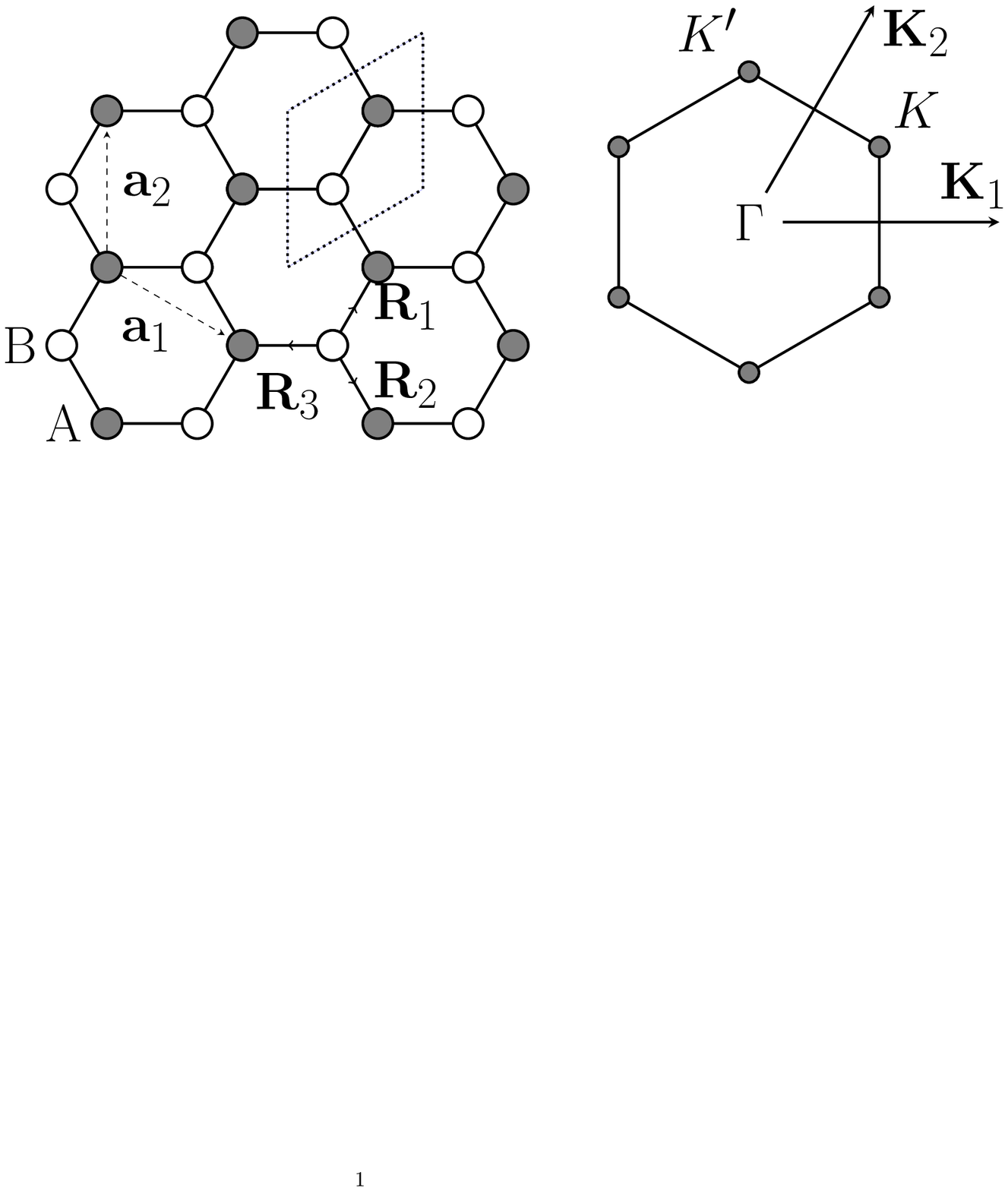}}  
\caption{The asymmetric hexagonal lattice in real space (on the left) and reciprocal space (on the right). The real space lattice vectors are ${\bf a}_1=a (\sqrt{3}/2,-1/2)$ and ${\bf a}_2=a (0,1)$, where $a$ is the lattice spacing (and for this geometry, $a=2\lambda/3$). The sublattices are connected by ${\bf R}_1=a(1/2\sqrt{3}, 1/2)$, ${\bf R}_2=a(1/2\sqrt{3}, -1/2)$, and ${\bf R}_3=a(-1/\sqrt{3}, 0)$. The dotted lines indicate the unit cell. The reciprocal lattice vectors are then ${\bf K}_1=(4\pi/\sqrt{3}a)(1,0)$ and ${\bf K}_2=(4\pi/\sqrt{3}a)(1/2,\sqrt{3}/2)$.} \label{fig:hex}
\end{figure}      

The honeycomb lattice can be viewed as two interpenetrating triangular sublattices, for $A$ and $B$, each with one site per unit cell (Fig. \ref{fig:hex}). 
With on-site energies of $\pm W$ on $A/B$ sites, and including only nearest neighbor hoppings, the Hamiltonian is
\begin{equation}
H({\bf k}) = \left( \begin{array}{cc} W & V({\bf k}) \\V^*({\bf k})& -W \end{array} \right) ,
\end{equation}
where $V({\bf k}) = -J[ e^{i{\bf k} \cdot {\bf R}_1} +e^{i{\bf k} \cdot {\bf R}_2}+e^{i{\bf k} \cdot {\bf R}_3} ]$. The two energy bands are then
\begin{equation} 
\varepsilon({\bf k}) = \pm \sqrt{W^2+|V({\bf k })|^2}\,.
\end{equation}

\begin{figure} [htdp]
\centering
$
\begin{array}{cc}
(a)\resizebox{0.26\textwidth}{!}{\includegraphics*{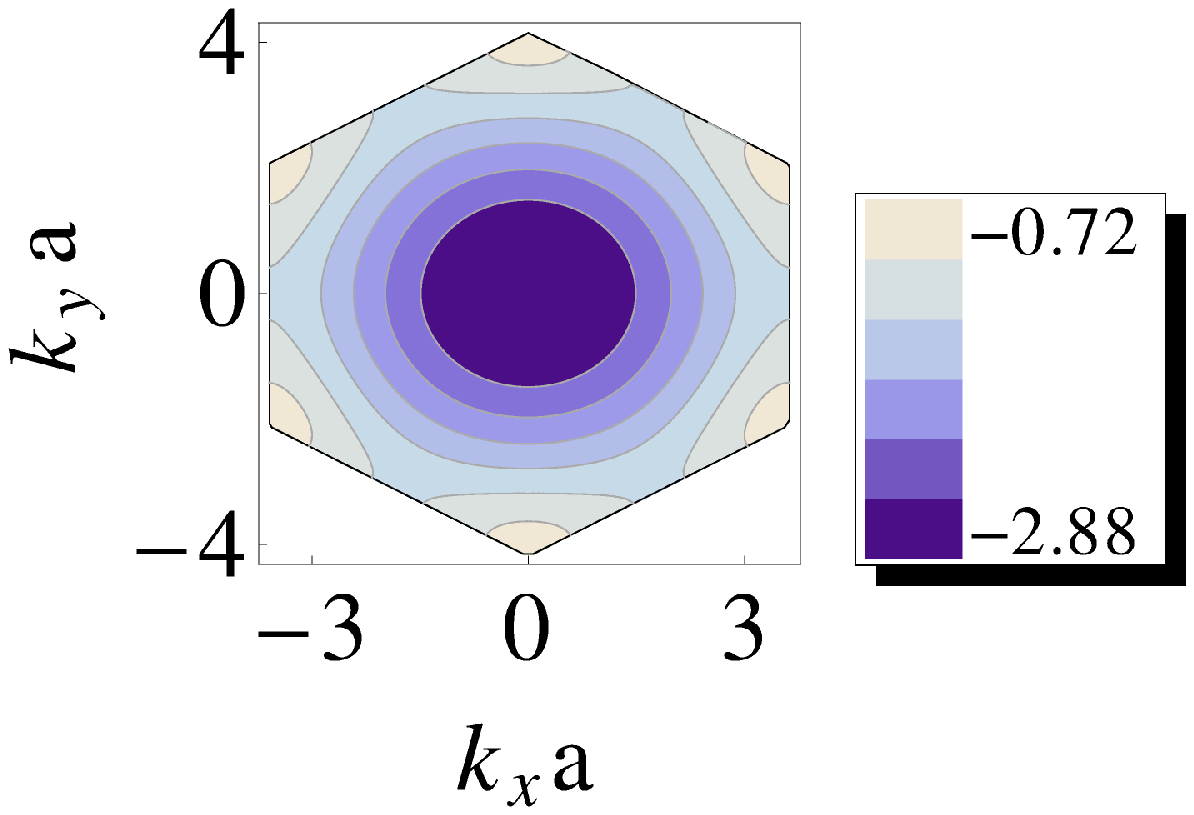}} &
(b)\resizebox{0.17\textwidth}{!}{\includegraphics*{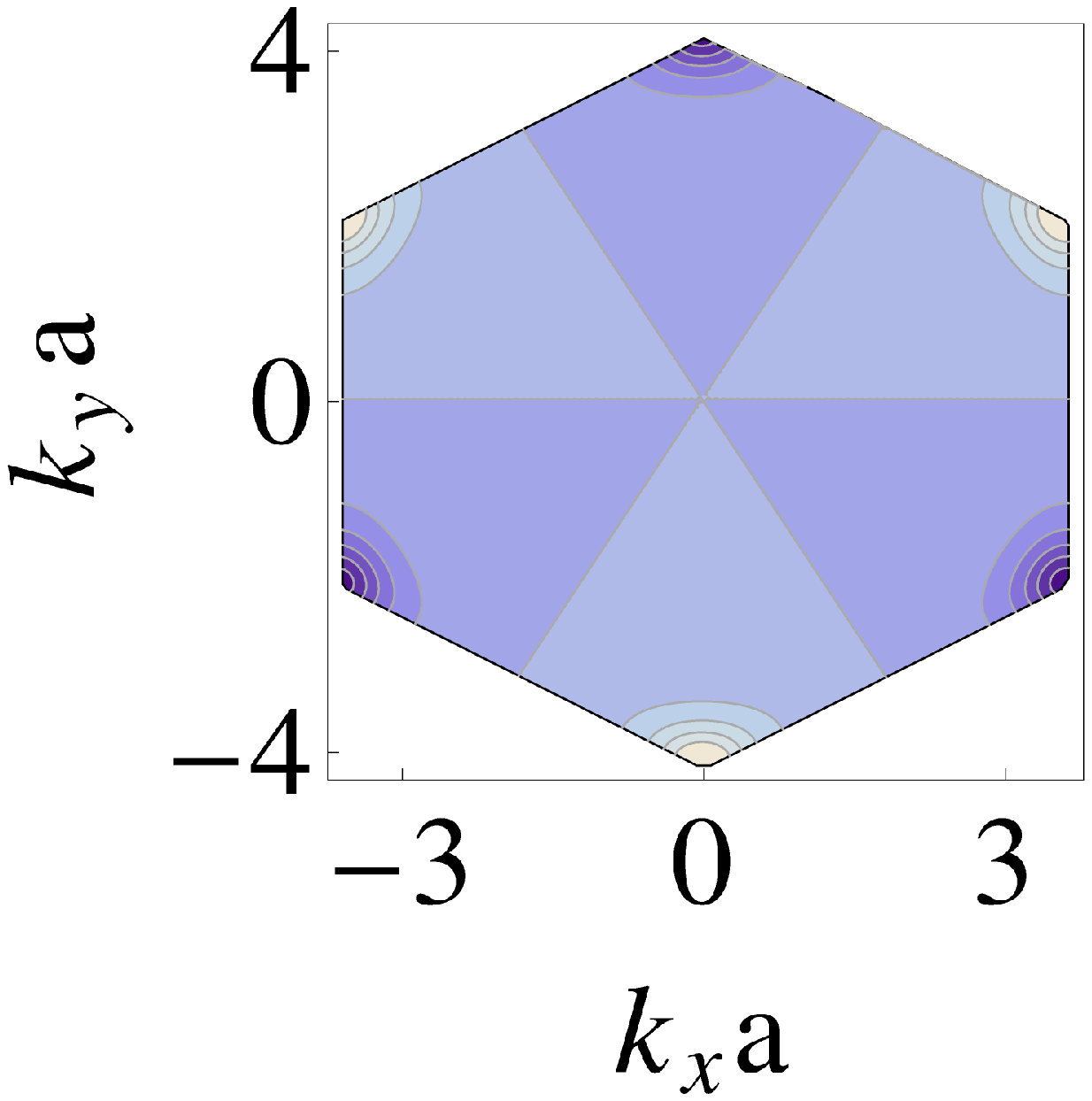}}  
\end{array}$
\caption{(a) Band structure of the lowest band of the asymmetric hexagonal lattice for $W=0.5 E_R$ and $t=1.0E_R$, and the energy in units of $E_R$ (for this geometry, $a=2\lambda/3$). Due to the asymmetry, gaps have opened at the Dirac points at the corners of the Brillouin zone. (b) The Berry curvature mapped out for $W=0.5 E_R$, using the method of Ref.~\onlinecite{fukui}. Light shading indicates $\Omega > 0$ and dark shading indicates $\Omega <0$. Positive and negative regions cancel, giving a net Chern number of zero.} \label{fig:hexchern}
\end{figure}

For $W=0$, the energy bands have two Dirac points at which $|V({\bf k})|=0$: these are at ${\bf k}=(4\pi/\sqrt{3}a)(1/2,1/2\sqrt{3})$ and ${\bf k}=(4\pi/\sqrt{3}a)(0,1/\sqrt{3})$, which we label as $K$ and $K'$. Near each of the Dirac points, the effective Hamiltonian takes a simple form. Close to the Dirac point $K$, writing ${\bf k} = (4\pi/\sqrt{3}a)(1/2,1/2\sqrt{3}) + {\bf q}$, the effective Hamiltonian is:
\begin{equation}
H({\bf q}) = \left( \begin{array}{cc} W & -\hbar v_F(q_x-iq_y) \\-\hbar v_F(q_x+iq_y)& -W \end{array} \right),
\end{equation}
where $\hbar v_F= (\sqrt 3/2) aJ$: a Dirac equation with mass. The band structure is shown in Fig. \ref{fig:hexchern}(a) for $W=0.5 E_R$ and $t=1.0E_R$. For this value of $W$, the band gap at the Dirac points is $1.0 E_R$ and the Landau-Zener tunneling probability is less than 0.09 for $|{\bf F}| = 1F_R$. Near the Dirac point, the Berry curvature is \cite{di}
\begin{equation}
\Omega({\bf q})= \frac{\hbar^2 v_F^2 W}{2(W^2+\hbar^2 v_F^2 q^2)^{3/2}}
\end{equation}
For the Dirac point $K'$,  writing ${\bf k} = (4\pi/\sqrt{3}a)(0,1/\sqrt{3}) + {\bf q}$, the Berry curvature has the same form but opposite sign. 

The resulting map of Berry curvature for the asymmetric hexagonal lattice is displayed in Fig. \ref{fig:hexchern}(b). This was previously found analytically in Ref.~\onlinecite{fuchs}. The Berry curvature around points $K$ and $K'$ has opposite signs such that the net Chern number of the band is zero. This vanishing Chern number is required by the fact that the system is time-reversal invariant.
\begin{figure} [htdp]
\centering
$
\begin{array}{cc}
(a)\resizebox{0.19\textwidth}{!}{\includegraphics*{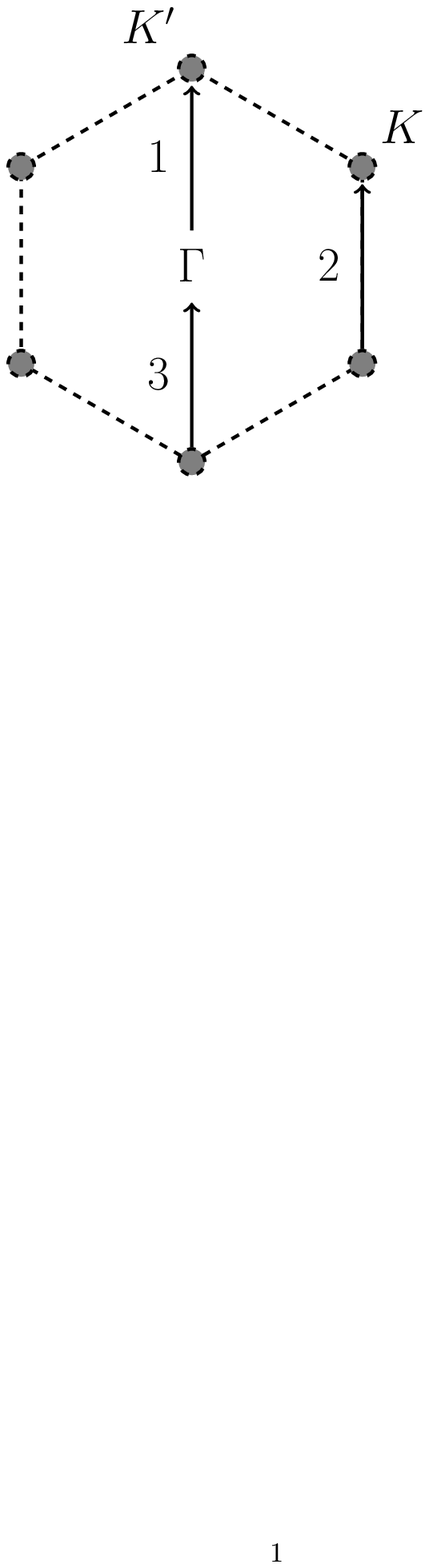}} &
(b)\resizebox{0.12\textwidth}{!}{\includegraphics*{dynlowforce}}  
\end{array}$
\caption{(a) The trajectory of a semiclassical wave packet through the Brillouin zone, starting from ${\bf k}=(0,0)$ with ${\bf F}=(0.0, 1.0F_R)$. The numbers indicate the order in which the path is traveled. (b) The corresponding real-space trajectory of the wave packet, starting from the origin, for $W=0.5 E_R$. This result was previously obtained in Ref.~\onlinecite{diener}. The motion along $x$ is due to the Berry curvature, while that along $y$ is due to the band structure. } \label{fig:dynhex}
\end{figure}      

\begin{figure} [htdp]
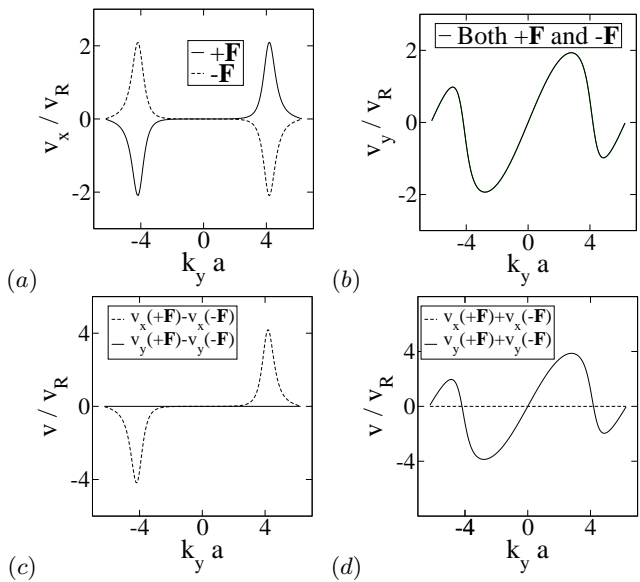

\centering
$
\begin{array}{cc}
(a)\resizebox{0.205\textwidth}{!}{\includegraphics*{vx}} &
(b)\resizebox{0.205\textwidth}{!}{\includegraphics*{vyhex}} \\
(c)\resizebox{0.205\textwidth}{!}{\includegraphics*{vchernhex}} &
(d)\resizebox{0.205\textwidth}{!}{\includegraphics*{vblochhex}} 
\end{array}$
\caption{(a) Velocity along $x$ of a wave packet moving under the force ${\bf F}=\pm(0.0, 1.0 F_R)$ with $W=0.5 E_R$. $k_y$ is measured along the path travelled (which is periodic in $4\pi$). (b) Velocity along $y$. (c) Applying the time-reversal protocol to extract the Berry velocity: ${\bf v}(+{\bf F})-{\bf v}(-{\bf F}) =  -2/\hbar ({\bf F} \times \hat{{\bf z}}) \Omega({\bf k})$. (d) Applying the time-reversal protocol to extract the group velocity: ${\bf v}(+{\bf F})+{\bf v}(-{\bf F}) = (2/\hbar) \partial \varepsilon({\bf k})/\partial {\bf k}$}. \label{fig:hexvel}
\end{figure}

From the Berry curvature and band structure, we can now calculate the semiclassical motion of a wave packet in this system (Fig. \ref{fig:dynhex}). To illustrate clearly the effects of Berry curvature, we start the wave packet at ${\bf k}=(0,0)$ and consider a force aligned along the $y$ direction, so that the 2D Bloch oscillation is also simply directed along $y$. This real-space trajectory was previously obtained in Ref.~\onlinecite{diener}, where the effects of a perturbing ``magnetic" field were also discussed. 

The velocities along $x$ and $y$ are shown in Fig. \ref{fig:hexvel}. As the wave packet passes through $K'$, the negative Berry curvature gives it a positive velocity in the $x$ direction. In between $K'$ and $K$, there is no curvature and it moves with a group velocity along $y$. When it passes through $K$, the positive curvature gives it negative $x$ velocity. As the regions of curvature have t he same magnitude, the effects cancel and there is no net drift. 

For ${\bf F}=(0,F)$ it is simple to determine the Berry curvature because the group and Berry velocities are perpendicular. For more general directions of the force, Lissajous-like oscillations will make it difficult to extract any information about the Berry curvature from the real-space motion. 

As proposed above, the Berry curvature may be cleanly mapped from the velocities using a time-reversal protocol. This is illustrated in Figs. \ref{fig:hexvel}(c) and \ref{fig:hexvel}(d). Here the velocities along $x$ and $y$ for $+{\bf F}$ and $-{\bf F}$ are combined to show the Berry velocity and the group velocity, respectively. 
   
\subsection{Optical flux lattices} \label{sec:ofl}

One of our main motivations for mapping the Berry curvature is to find a way of experimentally characterizing optical flux lattices. These have lately been proposed as schemes to access fractional quantum Hall physics in ultracold gases \cite{nigel, nigelnew}. The optical flux lattices have bands with nonzero Chern numbers. They consist of a state dependent potential in register with an interspecies coupling, and lead to effective magnetic flux with a high nonzero average (with a nonzero integer number of  magnetic flux quanta per unit cell). The specific optical coupling depends on the geometry and atomic species chosen. For atoms with a ground state and long-lived metastable excited state, such as an alkaline earth atom or ytterbium, a simple one-photon coupling scheme can be implemented \cite{nigel}. For more commonly used atomic species, such as alkali atoms, two hyperfine states can be used with coupling via two photon processes \cite{nigelnew}. Here, we discuss an example of each: the square one-photon optical flux lattice and the $F=1/2$ two-photon optical flux lattice. The lowest energy band in both lattices can have a Chern number of one and hence is topologically equivalent to the lowest Landau level, allowing quantum Hall physics to be accessed.

\subsubsection{One-photon square optical flux lattice} \label{sec:sq}

In this scheme, the electronic ground state and a long-lived excited state are coupled via a single photon process\cite{gerbier}. The Hamiltonian in the rotating wave approximation is
\begin{equation} \label{eq:Ha}
\hat{H}=\frac{{{\bf p}}^2}{2m} \hat{1}+  \hat{V}({\bf r}) ,
\end{equation}  
where $ \hat{1}$ and $\hat{V}$ are $2\times 2$ matrices acting on the two internal states of the atom. We neglect interactions, an assumption that is discussed further in Sec. \ref{sec:ex}. The square optical flux lattice is generated when \cite{nigel}
\begin{equation} \label{eq:M}
\hat{V}_{sq} = V[\cos(\kappa x) \hat{\sigma}_x + \cos(\kappa y) \hat{\sigma}_y +\sin(\kappa x)\sin(\kappa y) \hat{\sigma}_z ],
\end{equation}  
where $V$ sets the energy scale of the potential, $\hat{\sigma_{i}}$ are the Pauli matrices, $\kappa= 2 \pi /a$, and the lattice vectors are ${\bf a}_1=(a,0)$, ${\bf a}_2=(0,a)$. The flux density is everywhere of the same sign, and leads to a total flux per unit cell of $N_\phi = 2$ \cite{nigel}.

\begin{figure} [htdp]
\centering
$
\begin{array}{ccc}
(a)\resizebox{0.13\textwidth}{!}{\includegraphics*{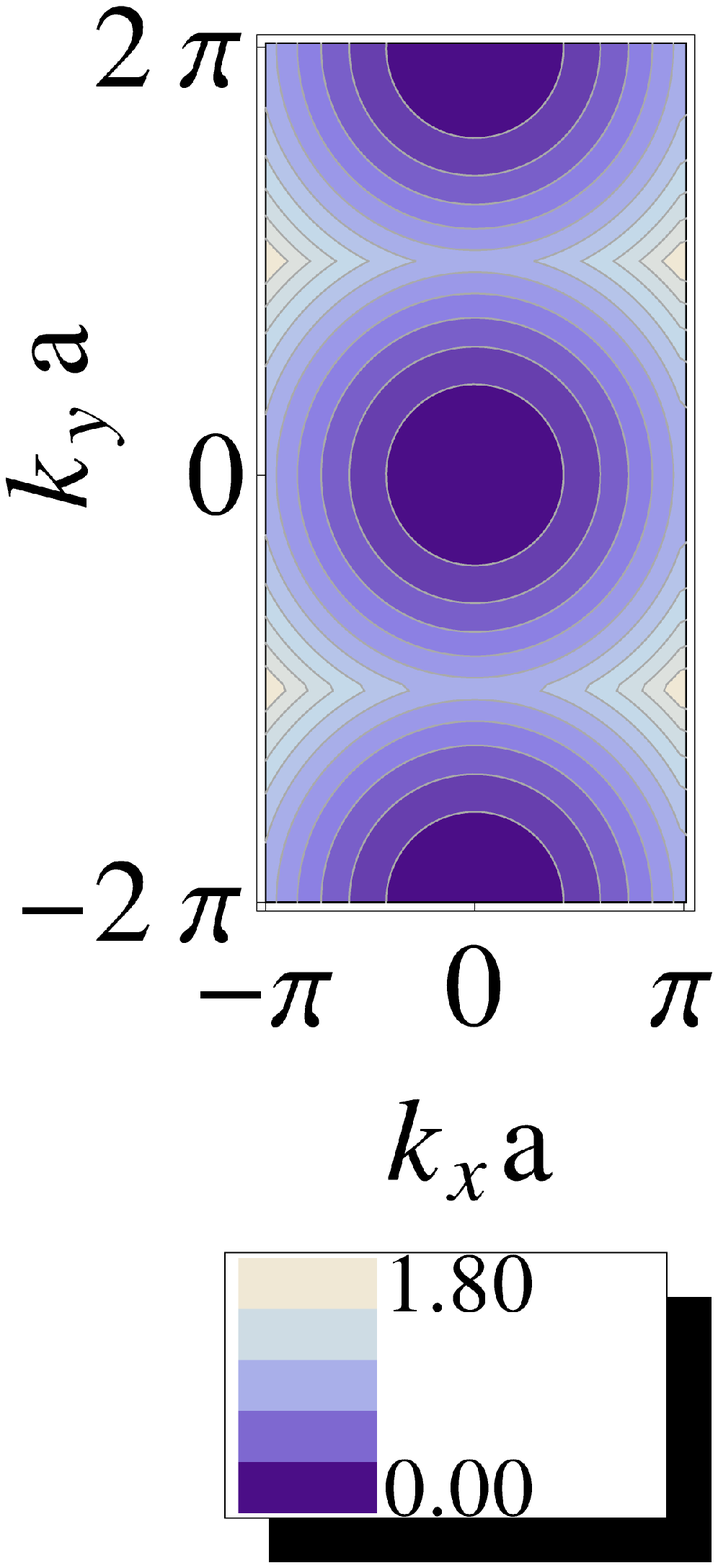}} & 
(b)\resizebox{0.13\textwidth}{!}{\includegraphics*{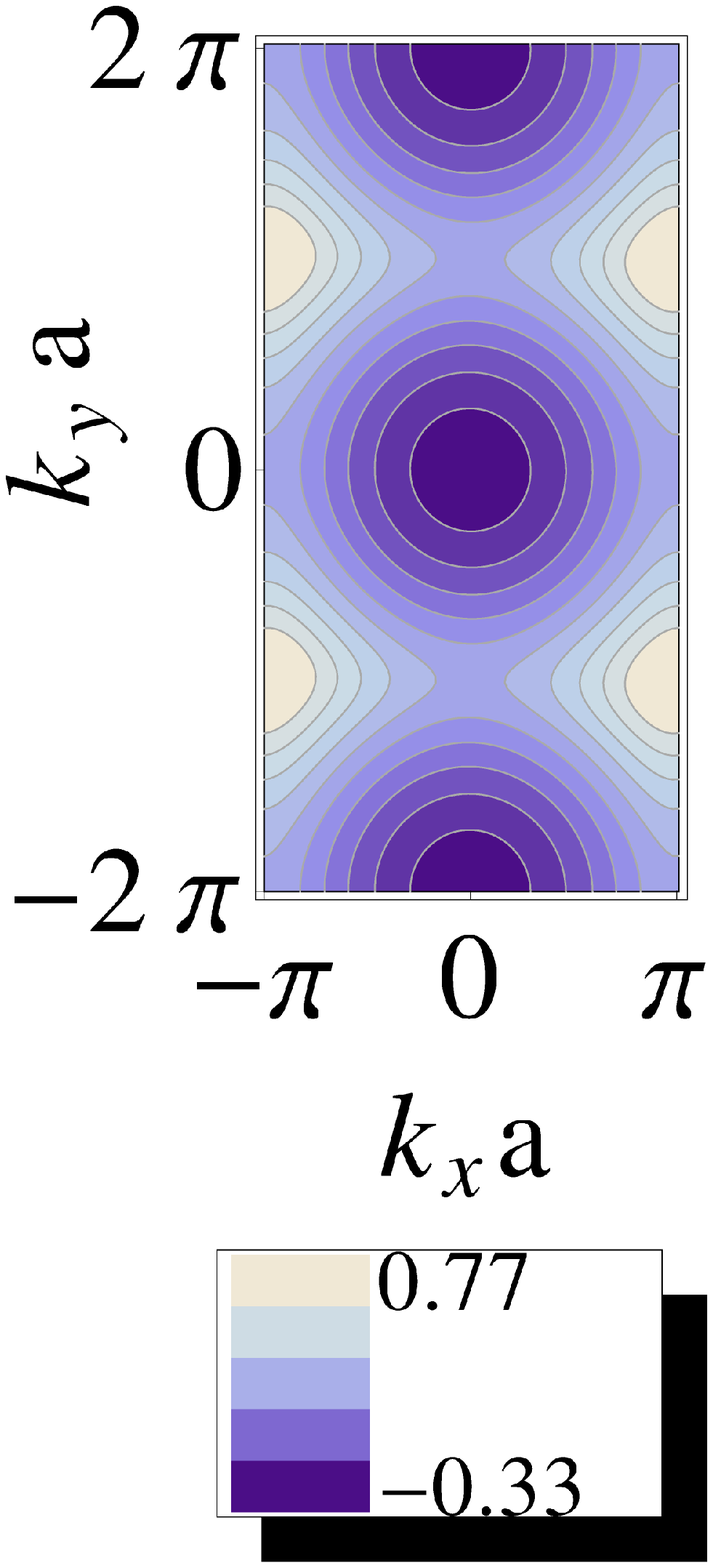}} & 
(c)\resizebox{0.13\textwidth}{!}{\includegraphics*{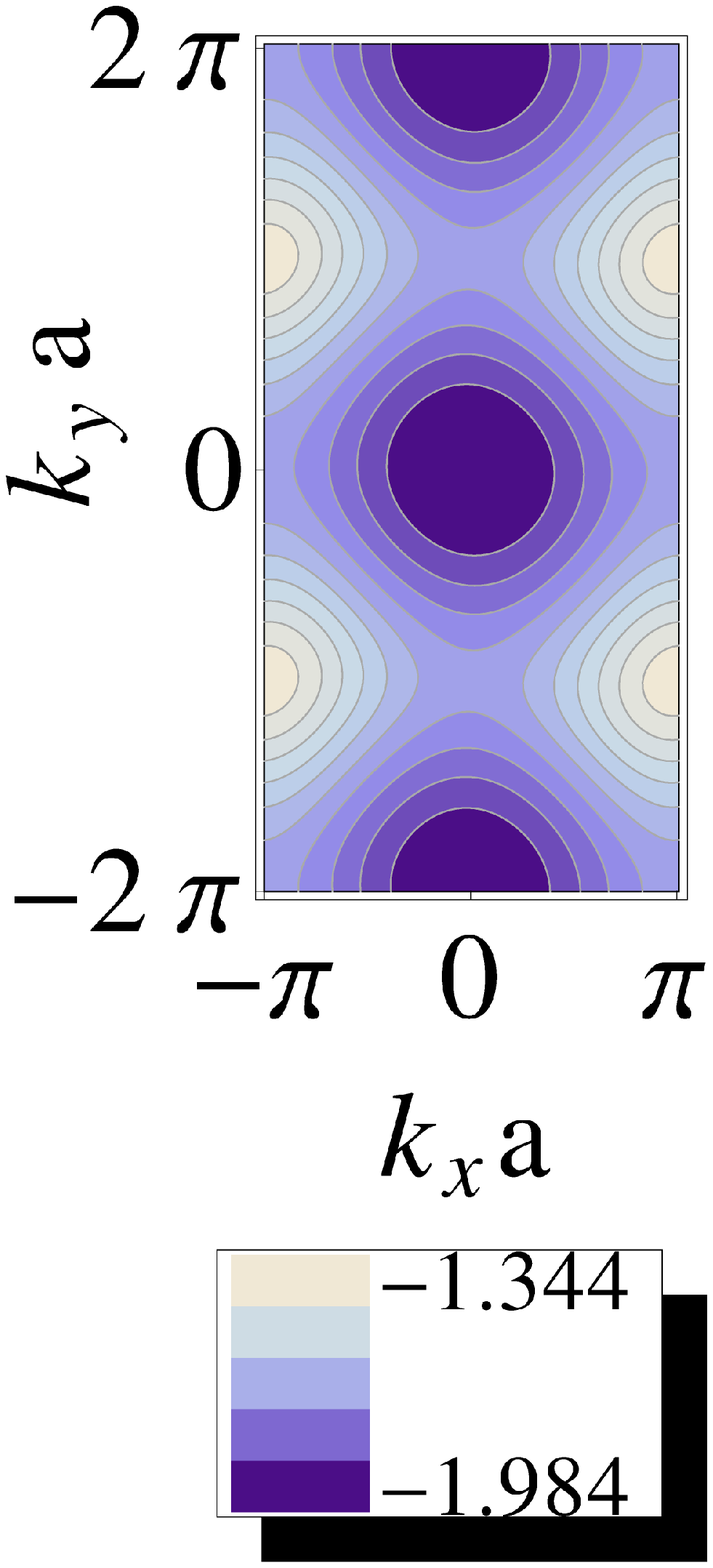}}  
\end{array}$
\caption{Band structure of the lowest band in the magnetic Brillouin zone for the one-photon square optical flux lattice with (a) $V=0.0E_R$, (b) $V=1.0E_R$ and (c) $V=3.0E_R$, and the energy in units of $E_R$.  At $V=0.0E_R$, the lowest band touches the second lowest band at $k=(\pi/a, \pm \pi/a)$. As $V$ is turned on, band gaps open at these points and Berry curvature is formed.}
\label{fig:sqbands}
\end{figure}      

We consider the regime where $V \lesssim \hbar^2 \kappa^2 /2 m = 4 E_R$ (for this geometry $a=\lambda/2$). We expand the periodic Bloch functions over a set of reciprocal lattice vectors ${\bf K}$,
\begin{equation}  \label{eq:bloch}
u_{n,{\bf k}}({\bf r})= \frac{1}{\sqrt{N}}\sum_{{\bf K}}e^{-i {\bf K} \cdot {\bf r}} \left( \begin{array}{c}
  c_{{\bf K}}^{1 (n,{\bf k})} \\  c_{{\bf K}}^{2 (n,{\bf k})}\\  \end{array} \right) 
\end{equation}       
and diagonalize the resulting Hamiltonian to find the band structure. As we consider low $V$, a small, finite set of ${\bf K}$ vectors will give the eigenfunctions and values to within the required numerical accuracy.  

The eigenfunctions are everywhere twofold degenerate, corresponding to two magnetic subbands. To distinguish between these states, we reinterpret the system within the magnetic Brillouin zone (MBZ) \cite{kohmoto, chang, nigel}. The optical coupling is invariant under the magnetic translation operators:
\begin{eqnarray}  \label{eq:v}
\hat{T}_{1} \equiv \hat{\sigma}_y e^{\frac{1}{2}{\bf a_1 \cdot \nabla}} & \qquad \qquad & \hat{T}_{2} \equiv \hat{\sigma}_x e^{\frac{1}{2}{\bf a_2 \cdot \nabla}} 
\end{eqnarray}  
which do not commute but satisfy $\hat{T}_{2}\hat{T}_{1} =-\hat{T}_{1}\hat{T}_{2} $. These operators represent rotations in spin space and translations by $\frac{1}{2}{\bf a}_{1,2}$, which enclose half a flux quantum (as $N_\phi = 2$) \cite{nigel, kohmoto}. The magnetic Brillouin zone is defined by a unit cell containing an integer number of flux \cite{kohmoto}; we choose a cell containing a single flux with vectors ${\bf a}_1$ and ${\bf a}_2 /2$. The corresponding commuting operators are $\hat{T}_{1}^2$ and $\hat{T}_{2}$, with eigenvalues $e^{i{\bf k}\cdot {\bf a}_1}$ and $e^{i{\bf k}\cdot {\bf a}_2/2}$. This defines the Bloch wavevector, ${\bf k}$, and the associated magnetic Brillouin zone \cite{nigel}. Now the first Brillouin zone covers $-\pi/a< k_x\leq \pi/a$ and $-2\pi/a < k_y\leq 2\pi/a $, doubling in size. Thanks to this unfolding, the lowest band is nondegenerate at each Bloch wavevector $\bf{k}$. The resulting band structure is shown in Fig. \ref{fig:sqbands} for $V=0.0E_R$, $V=1.0E_R$, and $V=3.0E_R$. 

\begin{figure} [htdp]
\centering
$
\begin{array}{cc}
(a)\includegraphics[width=0.22\textwidth]{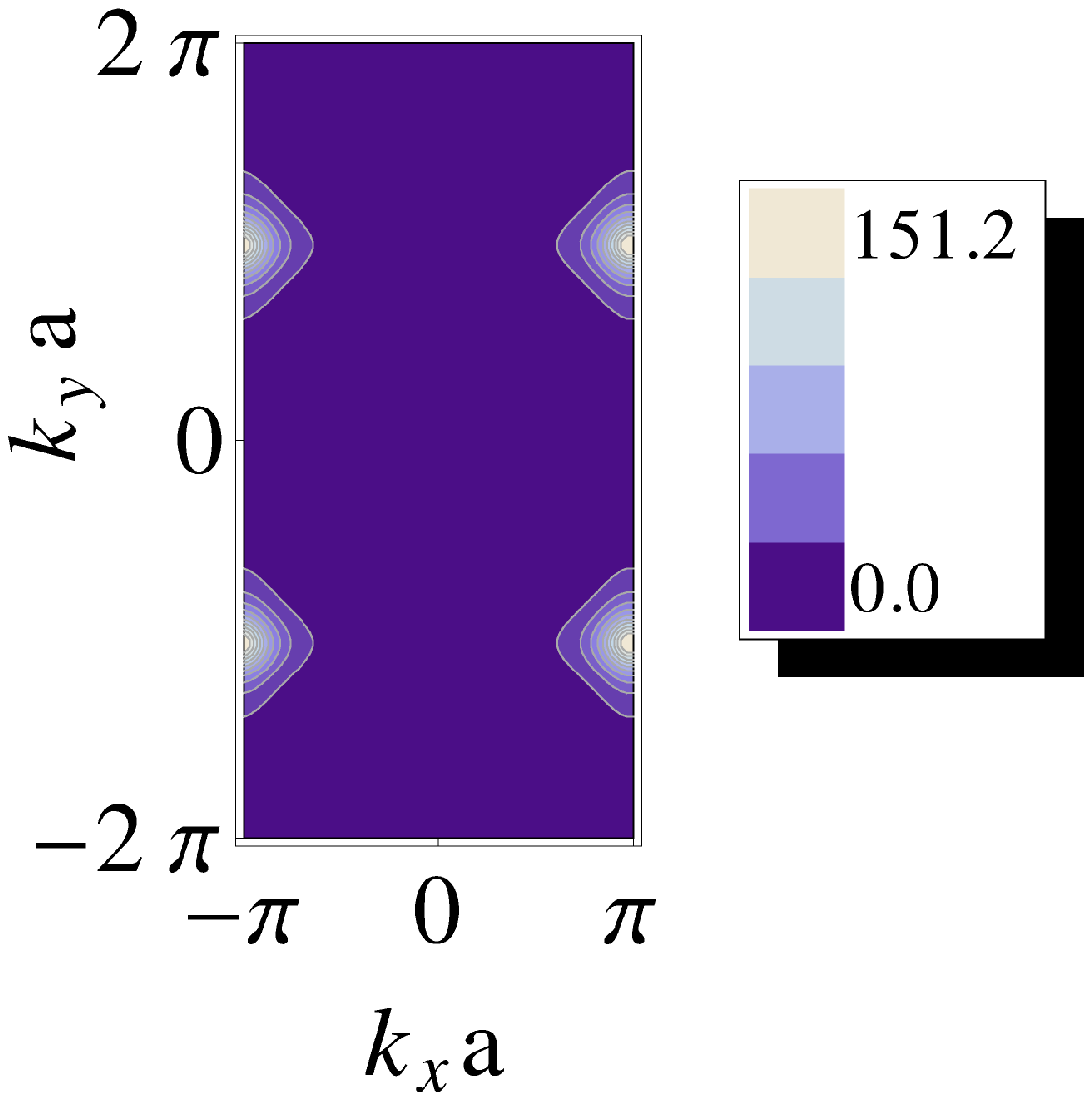} &
(b)\includegraphics[width=0.22\textwidth]{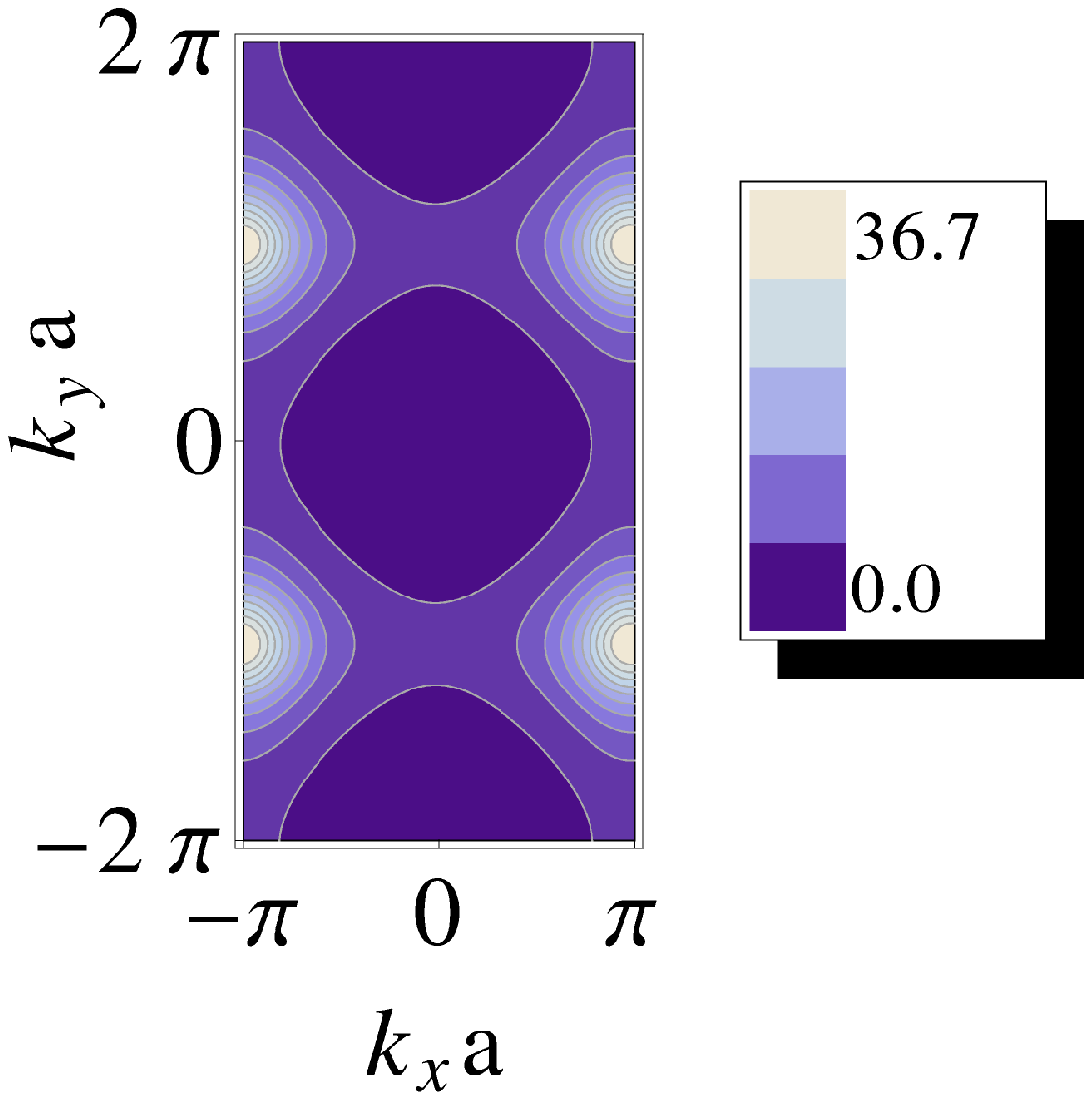} 
\end{array}$
\caption{Contour maps of the Berry curvature, $\Omega({\bf k})$, for the square optical flux lattice when (a) $V=1.0 E_R$ and (b) $V=3.0 E_R$. $\Omega$ is in units of $1/\mbox{(MBZ area)}=a^2/8\pi^2$. $\Omega$ is concentrated at points $k=(\pi/a, \pm \pi/a)$, spreading out with increasing $V$.} \label{fig:chernsq}
\end{figure}

The Berry curvature is shown over the magnetic Brillouin zone for the lowest band in Fig. \ref{fig:chernsq}, for $V=1.0 E_R$ and $V=3.0 E_R$. For nonzero $V$, the Chern number of this band is one, so it is analogous to the lowest Landau level. For small $V$ the Berry curvature, $\Omega$, is highly peaked at positions $k=(\pi/a, \pm \pi/a)$; as $V$ increases, $\Omega$ spreads out while remaining centered on these two points.  

\begin{figure} [htdp]
\centering
$
\begin{array}{cc}
(a)\resizebox{0.14\textwidth}{!}{\includegraphics*{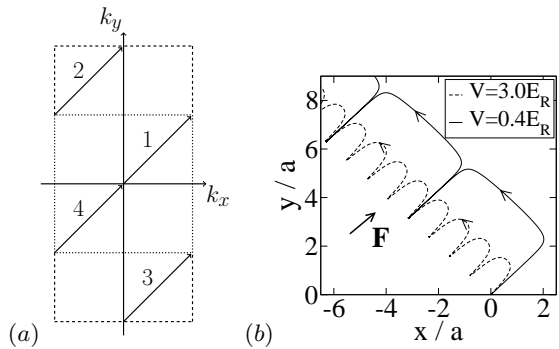}} &
(b)\resizebox{0.21\textwidth}{!}{\includegraphics*{dyn}}  
\end{array}$
\caption{(a) The trajectory of a semiclassical wave packet through the magnetic Brillouin zone, starting from ${\bf k}=(0,0)$ with $|{\bf F}|=1.0 F_R$ directed along (1,1). The numbers indicate the order in which the path is traveled. The dotted line indicates the simple Brillouin zone, while the dashed lines shows the extension into the magnetic Brillouin zone. (b) The corresponding real-space trajectory of the wave packet starting from the origin, for $V=0.4 E_R$ and $V=3.0 E_R$. The motion perpendicular to ${\bf F}$ is due to the Berry curvature, while that parallel is due to the bandstructure. } \label{fig:dynsq}
\end{figure}

\begin{figure} [htdp]
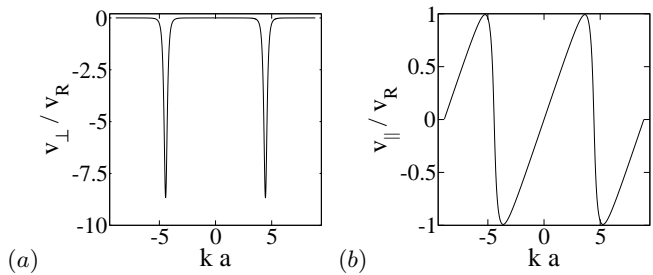

\centering
$
\begin{array}{cc}
(a)\resizebox{0.21\textwidth}{!}{\includegraphics*{velch}} &
(b)\resizebox{0.21\textwidth}{!}{\includegraphics*{velbl}}  
\end{array}$
\caption{The velocity of the wave packet (a) perpendicular and (b) parallel to the applied force for $|{\bf F}|=1.0 F_R$ directed along (1,1) and $V=0.4E_R$. $k$ is measured along the path travelled. For this simple alignment of the force, (a) contains the effects of Berry curvature along the path, while (b) shows the group velocity.} \label{fig:velsq}
\end{figure}

In Fig. \ref{fig:dynsq}(b) we illustrate the real-space trajectory for semiclassical motion of a wave packet in this optical flux lattice, for $V=0.4 E_R$ and $V=3.0 E_R$. As discussed above, we take $|{\bf F}|=1.0 F_R$, which is representative of the gravitational force $|{\bf F}|=m{\bf g}$ for $^{174}$Yb and $\lambda=578$ nm, parameters which are particularly relevant for this scheme \cite{nigel, gerbier}. As shown in Fig. \ref{fig:dynsq}(a), in momentum space the wave packet starts at ${\bf k}=(0,0)$ and moves under a force ${\bf F}$ parallel to the $(1,1)$ direction such that it passes through the points ${\bf k} = (\pi/a,\pm \pi/a)$ at which there is large positive Berry curvature. With ${\bf F}$ aligned along (1,1), the group velocity is parallel to the force, while the Berry velocity is perpendicular (Fig. \ref{fig:velsq}). As the Berry curvature is
everywhere positive, there is a net drift as successive regions of high $\Omega$ are traversed. For low potentials such as $V=0.4 E_R$ and $V=3.0 E_R$, the bandgap is small and the probability of Landau-Zener transitions on crossing the Brillouin zone boundaries is large. In an experiment where $|{\bf F}|=m {\bf g}$, this probability can be reduced below 0.1 by increasing the potential above $V=3.2E_R$. With higher $V$, the Berry curvature spreads out and hence the trajectory bends along more of its length. For clarity, we therefore discuss $V=0.4 E_R$ where the effects of Berry curvature and the group velocity are easiest to understand. The dynamics will be qualitatively the same for higher $V$, and we discuss the dependence of the motion on both $V$ and $|{\bf F}|$ in more detail in Sec. \ref{sec:ex}.  
 

As before, the simplicity of the trajectories and velocities relies on the alignment of ${\bf F}$ along a special direction. More generally in 2D, the Bloch velocity will not be parallel to the force and complex Lissajous-like figures will be observed; an example of this was previously shown in Fig. \ref{fig:lissa} for one oscillation at a low force, $|{\bf F}|=0.05 F_R$, where the effects of Berry curvature are small.  

\begin{figure} [htdp]
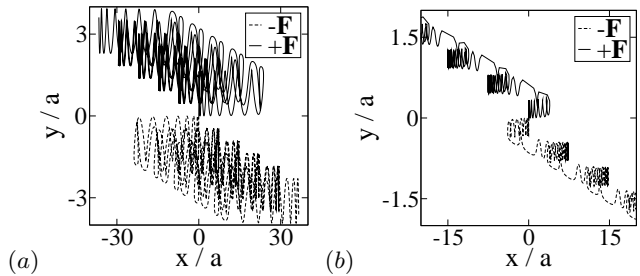

\centering
$
\begin{array}{cc}
(a)\resizebox{0.2\textwidth}{!}{\includegraphics*{dynlissalow}} &
(b)\resizebox{0.21\textwidth}{!}{\includegraphics*{lissadyn}} 
\end{array}$
\caption{Trajectory for a wave packet traveling with (a) $|{\bf F}|=1.0 F_R$ and (b) $|{\bf F}|=5.0 F_R$ where the force is aligned such that $F_x:F_y=1:16$ and $V=0.4E_R$. The Bloch motion is no longer purely along the direction of the force, and the resultant trajectory is complex. The net drift between Bloch oscillations is a measure of the total Berry curvature along a path, but other information is obscured. The trajectory due to the group velocity increases in size relative to the trajectory due to the Berry curvature with decreasing force (Sec. \ref{sec:ex}), leading to more complicated motion in (a) than (b).} \label{fig:lissasq}
\end{figure}

Here we demonstrate the effect of Lissajous-like figures on the motion for parameter ranges of interest. Figure \ref{fig:lissasq} shows the real-space trajectories for (a) $|{\bf F}|=1.0 F_R$ and (b) $5.0 F_R$, when the force is aligned such that $F_x:F_y=1:16$. The Bloch motion is no longer purely parallel to the force and while an average drift may be measured, the details of the motion due to Berry curvature have been lost. 

\begin{figure} [htdp]
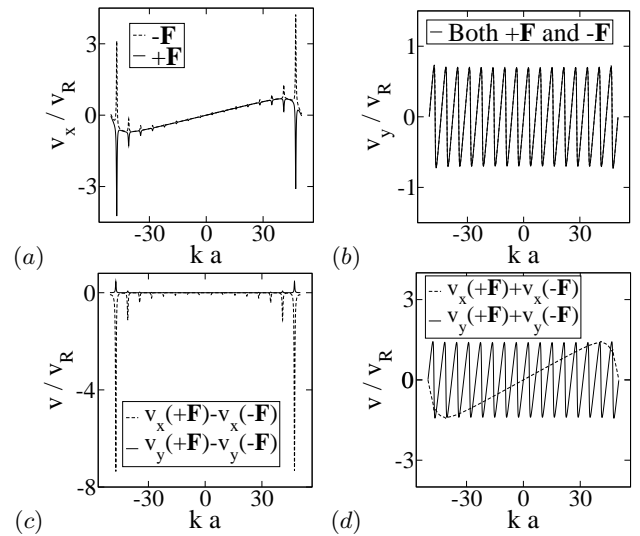

\centering
$
\begin{array}{cc}
(a)\resizebox{0.2\textwidth}{!}{\includegraphics*{vpplow}} &
(b)\resizebox{0.2\textwidth}{!}{\includegraphics*{vylow}} \\
(c)\resizebox{0.2\textwidth}{!}{\includegraphics*{vchernlow}} &
(d)\resizebox{0.2\textwidth}{!}{\includegraphics*{vblochlow}} 
\end{array}$
\caption{(a) Velocity along $x$ of a wave packet moving under the force $|{\bf F}|=1.0 F_R$ with $F_x:F_y=1:16$ and $V=0.4E_R$. $k$ is measured along the path traveled. (b)Velocity along $y$. (c) Applying the time-reversal protocol to extract the Berry velocity. The Berry velocity calculated from $v_x$ and $v_y$ differ by a factor of 16 from the ratio of the forces, and by a sign, due to the cross product. (d) Applying the time-reversal protocol to extract the group velocity. As expected, there are 16 oscillations from $v_y$ for every one from $v_x$.} \label{fig:vlow}
\end{figure}

\begin{figure} [h!]
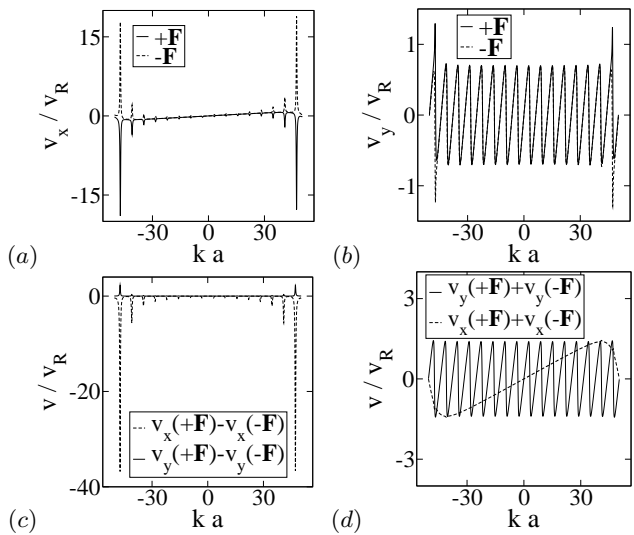

\centering
$
\begin{array}{cc}
(a)\resizebox{0.205\textwidth}{!}{\includegraphics*{vpp}} &
(b)\resizebox{0.20\textwidth}{!}{\includegraphics*{vy}} \\
(c)\resizebox{0.205\textwidth}{!}{\includegraphics*{vchern}} &
(d)\resizebox{0.20\textwidth}{!}{\includegraphics*{vbloch}} 
\end{array}$
\caption{As in Fig. \ref{fig:vlow} but now for $|{\bf F}|=5.0 F_R$.} \label{fig:vsq}
\end{figure}

\begin{figure} [htdp]
\centering
\resizebox{0.30\textwidth}{!}{\includegraphics*{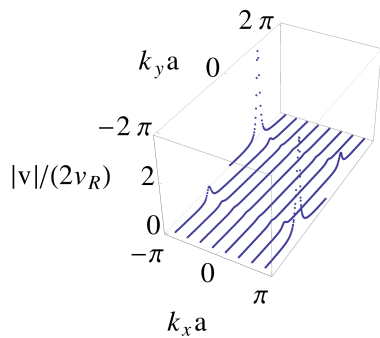}} 
\caption{$|v_x(+{\bf F})-v_x(-{\bf F})|/2=1/\hbar (F_y\Omega)$ plotted along the path taken by the wave packet in the Brillouin zone for $|{\bf F}|=1.0 F_R$ with $F_x:F_y=1:16$ and $V=0.4E_R$. This technique will enable experiments to directly map out the Berry curvature.} \label{fig:chernpath}
\end{figure}

We illustrate how our time-reversal protocol may be used to extract the local Berry curvature. Figures \ref{fig:vlow} and \ref{fig:vsq} show the velocities along $x$ and $y$ over the path through the Brillouin zone. By comparing the time-reversed velocities, quantities proportional to the Berry velocity and the group velocity are extracted. Figure \ref{fig:chernpath} shows the resulting map of Berry curvature over the Brillouin zone for $|{\bf F}|=1.0 F_R$. The same result is obtained (up to a scale factor) from either of these two cases of different magnitude of the force. 

\subsubsection{Two-photon optical flux lattice for $F=1/2$}

To generate optical flux lattices for the more commonly used alkali
atoms one must employ dressed states involving a two-photon
  coupling \cite{nigelnew}. We consider the representative case of a
lattice with triangular symmetry where the hyperfine states coupled
have angular momentum $F=1/2$, as for $^{171}$Yb.
Qualitatively similar semiclassical dynamics are expected for other
values of $F$, such as the experimentally common case of $F=1$ for
$^{87}$Rb \cite{nigelnew}.  The two-photon optical flux lattice
 we study leads to a net effective magnetic field in real space in which there is
  one flux quantum per unit cell. Semiclassical motion within
  a similar scheme has previously been studied in
  Ref.~\onlinecite{dudarev}. In that scheme, the artificial magnetic field
  is still locally nonzero, but the flux per unit cell vanishes
  \cite{dudarev, nigelnew} and the Chern number of each band is
  zero. For the optical flux lattice described here, the 
  energy bands may have nonzero Chern numbers.

For the $F=1/2$ optical flux lattice, two hyperfine ground states, $g_\pm$, with angular momentum $J_g=1/2$, are coupled to an excited state, $e$, also with angular momentum $J_e=1/2$, via an off-resonance excitation that ensures the population of $e$ remains negligible. The Hamiltonian then acts in the $g_\pm$ manifold, with the form of (\ref{eq:Ha}). The details of the optical coupling and the resulting Hamiltonian are discussed in Ref. \cite{nigelnew} and in Appendix \ref{sec:appa}.

The final Hamiltonian can be written as:
\begin{equation} 
\hat{H}'= \hat{U}^\dagger \hat{H} \hat{U}= \frac{({\bf p}-\hat{\sigma}_z \hbar {\bf k}_3/2)^2}{2m} + \hat{V}' , \label{eq:couple}
\end{equation}  
where ${\bf k}_3= k (0, 1, 0)$, and $\hat{U}$ is a unitary transformation that is applied to expose the full  symmetry of the system. The transformed optical potential $\hat{V}'$ has the maximal translational symmetry, causing  $\hat{{\bf p}}/\hbar$ to be conserved up to the addition of the reciprocal lattice vectors ${\bf K}_{1/2} =-k/2 (\pm \sqrt 3, 3, 0)$ (Appendix \ref{sec:appa}). The resulting Brillouin zone, defined by ${\bf K}_{1/2}$, is equivalent to the asymmetric hexagonal lattice in Sec. \ref{sec:hex}. As can be seen, an important feature of the two-photon optical flux lattice is that, under this unitary transformation, the momenta of $g_\pm$ are offset by $\pm \hbar{\bf k}_3/2$. This offset does not affect the semiclassical equations of motion, which determine the rate of change of (crystal) momentum under an applied force (\ref{eq:motion}),  and which still apply for the bands formed from the eigenstates of $\hat{H}'$. 

The optical coupling, $\hat{V}'$, is characterized by parameters $\epsilon$ and $\theta$ as well as the overall strength of the potential $V$ (Appendix \ref{sec:appa}). When $\theta=\epsilon=0$, the optical potential does not couple the states $g_+$ and $g_-$, and acts on each simply as a scalar potential, with the same symmetries as the hexagonal lattice discussed above when inversion symmetry is unbroken. The offsets of the momenta of $g_\pm$ shift the Dirac points of these two states relative to each other.
A small nonzero $\theta$ breaks inversion symmetry and opens up gaps at the Dirac points in such a way that the pairs of bands are topologically trivial, i.e., have a net Chern number of zero. Nonzero $\epsilon$ and $\theta$ together break time-reversal symmetry and lead to bands with nonzero Chern number. 

\begin{figure} [htdp]
\centering
$
\begin{array}{cc}
(a)\resizebox{0.22\textwidth}{!}{\includegraphics*{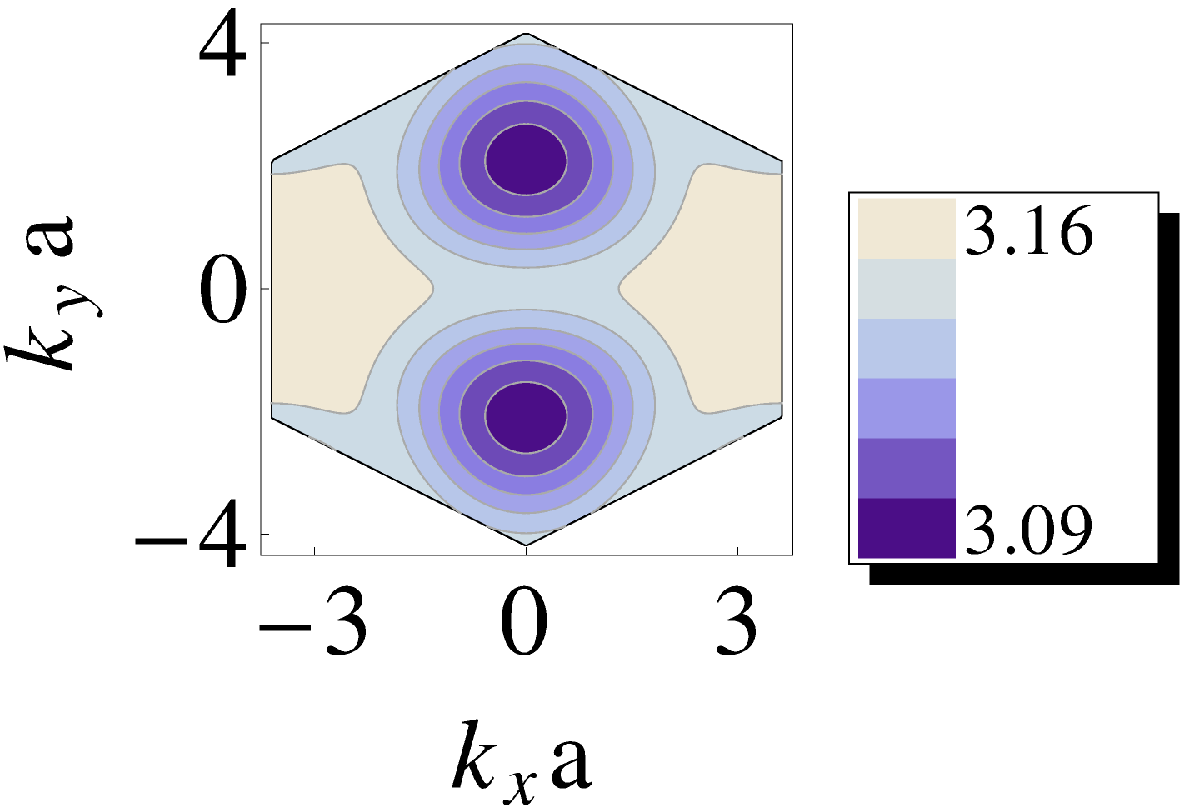}} &
(b)\resizebox{0.22\textwidth}{!}{\includegraphics*{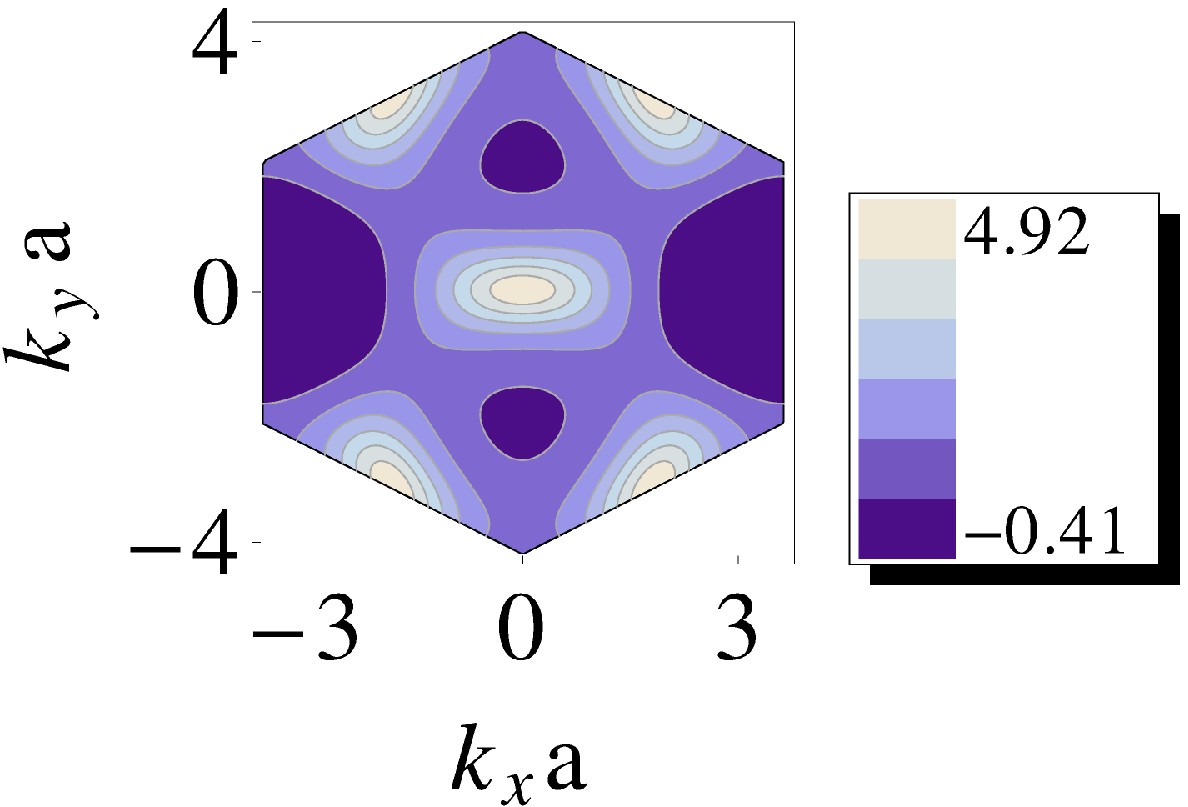}}  
\end{array}$
\caption{(a) The energy (measured in $E_R$) of the lowest band for $V=1.8 E_R$, $\theta=0.3$, and $\epsilon=0.4$. For these parameters, the lowest band has a Chern number of one. (b) The corresponding Berry curvature. $\Omega({\bf k})$ is in units of $1/\mbox{(BZ area)}$. In this case, the Berry curvature is significant over much of the Brillouin zone and has regions of both positive and negative sign.} \label{fig:alkbands}
\end{figure}   

\begin{figure} [htdp]
\centering
\resizebox{0.3\textwidth}{!}{\includegraphics*{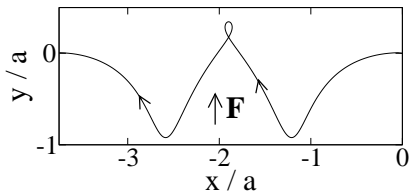}}  
\caption{The real space trajectory of a wave packet under a force ${\bf F}=(0.0, 0.1 F_R)$ over one period, for $V=1.8 E_R$, $\theta=0.3$ and $\epsilon=0.4$. The wave packet follows the path described in Fig. \ref{fig:dynhex}. While the group velocity is again purely along $y$, the motion due to the Berry curvature is more complicated due to the varying sign of $\Omega$.} \label{fig:alkdyn}
\end{figure} 

We have numerically calculated the band structure and Berry curvature of the lowest band for a case of large coupling, when $V=1.8 E_R$, $\theta=0.3$, and $\epsilon=0.4$, for which this band has a Chern number of one (Fig. \ref{fig:alkbands}). Note that in this case, the energy minimum of the band is not at ${\bf k}=(0,0)$. Applying the force along $(0,1)$, the wave packet will follow the same path as in the asymmetric hexagonal lattice (Fig. \ref{fig:dynhex}). We choose $|{\bf F}|=0.1F_R$ to ensure that the Landau-Zener tunneling probability is small (below 0.1). The resulting real-space trajectory is shown in Fig. \ref{fig:alkdyn}. Due to the simple alignment of the force, we can again associate the $y$ motion with the group velocity and the $x$ motion with the effects of the Berry curvature. 

As the Berry curvature is substantially spread out through the Brillouin zone (Fig. \ref{fig:alkbands}), there is now a continual drift along $x$ over the trajectory. Since the Chern number is one, the Berry curvature is (largely) of the same sign along the trajectory, leading to a net drift of the wave packet over one period of the Bloch oscillation. In this lattice, there are also regions of both positive and negative Berry curvature. When the Berry curvature is positive, the wave packet travels in the negative $x$ direction, while when $\Omega$ is negative, the wave packet moves along the positive $x$ axis. The sign change will therefore not be detected if only the total drift is measured. Instead, the time-reversal procedure described above can again be applied to cleanly map out the local Berry curvature.   
 
  \section{Experimental Considerations} \label{sec:ex}

We now consider how the velocity may be measured and how feasible it will be to observe the Berry curvature effects experimentally. 

The mean atomic velocity may be extracted from the time-of-flight expansion image. This measures the momentum distribution \cite{blochdalibardzwerger}, from which the mean momentum $\langle {\bf p}\rangle$ may be deduced by the weighted average. The mean atomic velocity of the initial wave packet then follows from Ehrenfest's theorem as $\langle {\bf v}\rangle = \langle {\bf p}\rangle/m$. This approach was successfully experimentally applied by Ben Dahan {\it et al.} \cite{dahan} to detect Bloch oscillations in a one-dimensional lattice. The same approach applies for dressed states of internal atomic states, governed by (\ref{eq:Ha}); in this case the mean velocity can be obtained from the average momentum over all internal states.

Alternatively, it is possible to extract the velocity directly from measurements of the center of mass motion in real space. Thanks to recent experimental advances, the position of the wave packet's center of mass may be imaged with a high resolution, on the order of a lattice spacing \cite{esslingerprivate}. For bands with nonzero Chern number, the Berry curvature can cause the wave packet to have a net drift in space over each period of the Bloch oscillation; this leads to large cumulative effects on the position of the wave packet over many oscillation periods. As described in Sec. \ref{sec:relchern}, measurements of the position of the wave packet therefore offer a sensitive way to show that the Chern number of the band is nonzero. Indeed measurements of these drifting trajectories in constant applied force are equivalent to measurements of the {\it edge states} that must arise for a band with nonzero Chern number.

To this end, we consider how to maximize the importance of the effects of the Berry curvature relative to those of the group velocity. In the cases described above where the bands almost touch, we can consider the band gap, $\delta \varepsilon$, as a small parameter. This applies to the hexagonal lattice when asymmetry is small and to the square one-photon optical flux lattice for small potential, $V$. (Note that the two-photon optical flux lattice is considered far from the band closing regime.) In the small band gap limit, the momentum width $\delta k$, over which the band is changed will be $\delta k\simeq \delta \varepsilon/(\hbar v_R)$, where $v_R$ is the typical group velocity at the zone boundary for vanishing band gap. The Berry curvature is therefore nonzero over the area, $A \simeq (\delta k )^2$. Assuming the Berry curvature is uniform within this, the invariance of the Chern number means $\Omega \simeq 1/A \simeq 1/(\delta k)^2$. The Berry velocity is ${\bf v}_\Omega =  (\dot{{\bf k}}\times \hat{{\bf z}}) \Omega$ so, as the wave packet traverses this region in one Bloch oscillation, the Berry curvature leads to a displacement of size $\int v_\Omega dt \simeq \delta k \Omega \simeq 1/\delta k
\simeq \hbar v_R/\delta \varepsilon$.
 Conversely, the typical group velocity is $\simeq v_R$ so over
   the period of one Bloch oscillation, $\tau_B \simeq  h / (F a)$, the 
typical amplitude of displacement is $v_R \tau_B \simeq h v_R/(F a)$.
Thus, these contributions to the real-space trajectory have different
dependences on $\delta \varepsilon$: the Berry curvature contribution scales
as $1/\delta\varepsilon$, while the contribution from the group velocity is
independent of $\delta \varepsilon$.  This is found in our numerical
results, but is shown only qualitatively in the results presented in Fig.
\ref{fig:dynsq}, as in this case $\delta \varepsilon$ is not that
  small. The effects of Berry curvature can therefore be maximized with
  respect to those of the group velocity by choosing a small band gap. 
Note also that the two contributions have different dependences on the size of the force:
the displacement due to the Berry curvature is independent of force, while that due to the group velocity is inversely proportional to it. Therefore, the effects of Berry curvature will be most evident for a high external force. Note, however, that there are some practical limitations on both the choice of force and band gap. To ensure that the evolution of the wave packet is adiabatic, the rate of Landau-Zener tunneling to the next lowest band should be negligible [Eq. \ref{eq:zener}]. The assumption of adiabatic evolution is therefore violated when the applied force is too high and the band gap is too small. Also, for small $\delta \varepsilon$, when the bands nearly touch, the Berry curvature becomes concentrated in small regions, of area $(\delta k)^2$. This then requires the momentum of the wave packet and the alignment of the force to be precisely controlled in order to direct the wave packet through this region. For intermediate band gaps, the curvature is spread out. A natural compromise is to choose the band gap such that $\delta k$ is as small as the momentum uncertainty $1/w$ with which a wave packet can be prepared (here $w$ is the spatial width of the wave packet); one then expects the displacement due to the Berry curvature to be $1/\delta k\simeq w$ on the order of the spatial extent of the wave packet.

From our numerical calculations, we can quantitatively estimate the lengthscales of the dynamics. For example, we consider the dynamics of $^{174}$Yb atoms in the one-photon optical flux lattice, with ${\bf F}=m{\bf g}$ along (1,1), $\lambda=578$ nm, and $V=3.2 E_R$. For this choice of parameters, the Landau-Zener tunneling probability given by (\ref{eq:zener}) is approximately 0.1. The wave packet follows the Brillouin zone path in Fig. \ref{fig:dynsq}(a) and has a real-space trajectory similar to that of the dashed line in Fig. \ref{fig:dynsq}(b). As the wave packet moves along section 1 of its path [from ${\bf k}=(0,0)$ to ${\bf k}=(\pi/a,\pi/a)$], it moves with an average group velocity of 1.0 mm s$^{-1}$, traveling approximately 0.6 $\mu$m in the direction of the force in real space. For $V=3.2 E_R$, the Berry curvature is substantially spread out over the Brillouin zone, and so the wave packet's trajectory bends as it travels, moving it 0.3 $\mu$m perpendicular to the force.  As the wave packet continues from ${\bf k}=(-\pi/a,\pi/a)$ to ${\bf k}=(0,2\pi/a)$, the group velocity changes sign and the wave packet travels 0.6 $\mu$m in the opposite direction to the force. The Berry velocity does not change sign, and so the wave packet moves a further 0.3 $\mu$m perpendicular to the force. This behavior repeats for sections 3 and 4 of its path. The average Berry velocity over one complete oscillation is therefore approximately 0.3 mm s$^{-1}$. If the force is slightly misaligned, the trajectory will instead be a Lissajous-like oscillation, approximately bounded by a box of diagonal length 0.6 $\mu$m. For the same Berry velocity, the wave packet would then take approximately 2.0 ms to drift this distance. These length and time scales are within current experimental capabilities.   

\begin{figure} [htdp]
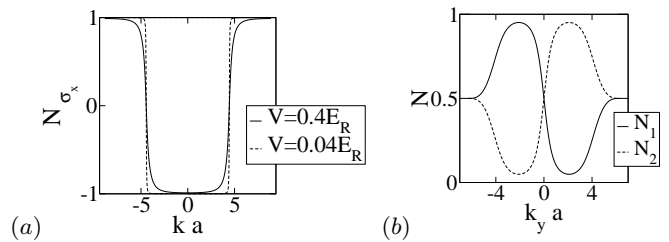

\centering
$
\begin{array}{cc}
(a)\resizebox{0.24\textwidth}{!}{\includegraphics*{oscill}} &
(b)\resizebox{0.185\textwidth}{!}{\includegraphics*{alkosc}}  
\end{array}$
\caption{(a) Oscillation in expectation value of $\hat{\sigma}_x$ for the one-photon optical flux lattice, as the wave packet travels along $(1,1)$. The dashed line is for $V=0.04E_R$, while the smooth line is $V=0.4E_R$. The area over which the transfer occurs decreases with potential in a similar way to how the Berry curvature area decreases. $k$ is measured along the path of the wave packet. (b) Oscillation in population in the two states for the two-photon optical flux lattice as the wave packet travels along $(0,1)$ for $V=1.8 E_R$, $\theta=0.3$, and $\epsilon=0.4$. The maximum population imbalance appears to be around points $K$ and $K'$. $k_y$ is measured along the path of the wave packet. } \label{fig:osc} 
\end{figure}

Two additional practical considerations are the effects of dispersion and interactions. The wave packet will spread as it travels, and this dispersion could obscure the dynamics described. However, provided the center of mass of the wave packet can be measured to an accuracy greater than its width, this should not prevent the observation of Berry curvature effects.

Interactions destroy the coherence of a wave packet over time and can have a strong dephasing effect on Bloch oscillations \cite{roati, smerzi, anderson}. Nonlinearity can also lead to the collapse of the wave packet into discrete solitons \cite{smerzi, solitons}. We have ignored the effects of interactions in our analysis, an approximation suitable over these time scales for fermionic atoms \cite{roati}, for species with low scattering lengths \cite{ferrari}, or where the interaction strength can be tuned to zero by means of a Feshbach resonance \cite{gustavsson, fattori}. 

One can also look for distinct features in the momentum distributions of atoms undergoing Bloch oscillations in the optical flux lattices discussed above (Fig. \ref{fig:osc}). In the two-photon optical flux lattice, the population of atoms oscillates between the two internal states as the wave packet moves through the Brillouin zone. Near the points $K$ and $K'$, the population imbalance is maximum but of opposite sign, reflecting the characteristics of the Berry curvature. In the one-photon optical flux lattice, the unfolding into the full magnetic Brillouin zone means the Bloch states are eigenstates of $\hat{\sigma}_x$. As a result, the oscillation takes place between those superpositions of the internal states that are eigenstates of  $\hat{\sigma}_x$. The transfer between eigenstates occurs over an area which decreases with potential, in a similar way to how the Berry curvature area decreases.  

\section{Conclusions}

In summary, we have proposed a general method for mapping the local Berry
curvature over the Brillouin zone in ultracold gas experiments. The Berry
curvature crucially modifies the semiclassical dynamics and so affects the
Bloch oscillations of a wave packet under a constant external force. In
particular, the Berry curvature may lead to a net drift of the wave packet with
time. However, two-dimensional Bloch oscillations are interesting in their own
right, and one may lose information about the Berry curvature due to the complicated Lissajous-like figures that may arise. 

We have shown that this information can be recovered via a time-reversal protocol. The group velocity at a point in the Brillouin zone is invariant under a reversal of force, while the Berry velocity changes sign. As a result, the velocities under positive and negative force can be compared to extract either one or the other. This protocol will allow the local Berry curvature to be cleanly mapped out over the path of the wave packet through the Brillouin zone. 

We have demonstrated this protocol using the semiclassical dynamics of three model systems which are currently of experimental interest: the asymmetric honeycomb lattice and two optical flux lattices. Finally, we have discussed various experimental considerations, including how the velocity may be measured and how to maximize the magnitude of the Berry curvature effects on the dynamics. These methods can be used to characterize the topological character of band structures of complex optical lattices including optical flux lattices. \\

\acknowledgments{We are grateful to Tilman Esslinger for helpful conversations and to Eugene Demler
for insightful comments on the importance of Ehrenfest's theorem.
This work was supported by EPSRC Grant EP/F032773/1.}

\appendix

\section{Hamiltonian for the Two-Photon Optical Flux Lattice for $F=1/2$} \label{sec:appa}

The Hamiltonian acts in the $g_\pm$ manifold, with the form of (\ref{eq:Ha}). The optical potential, $\hat{V}({\bf r})$, is characterized by the Rabi frequencies $\kappa_m$, $m=0, \pm 1$, where $m\hbar$ is the angular momentum along $z$ gained by the atom when it absorbs a photon \cite{nigelnew}. 

The potential is formed from laser beams at two frequencies: $\omega_L$ and $\omega_L+\delta$ where $\delta$ is the Zeeman splitting of the ground states. The laser beams at $\omega_L$ are linearly polarised and traveling in the $xy$ plane, while the laser at $\omega_L+\delta$ gives a plane wave propagating along the $z$ axis with $\sigma_z$ polarization. The resulting potential takes the form:
\begin{equation} 
\hat{V} = \frac{\hbar \kappa^2_{tot}}{3 \Delta} \hat{1} + \frac{\hbar}{3 \Delta}  \left( \begin{array}{cc} |\kappa_-|^2-|\kappa_+|^2 & E \kappa_0 \\E \kappa_0^* &  |\kappa_+|^2-|\kappa_-|^2 \end{array} \right)
\end{equation}  
where $\kappa_{tot}^2= \sum_m |\kappa_m|^2$, $\Delta=\omega_L-\omega_A$, with $\omega_A$ as the atomic resonance frequency and it is assumed that $|\Delta| \gg |\delta|, |\kappa_m|$. The field of the laser at frequency $\omega_L+\delta$ is characterized by $E$, which serves as a uniform, adjustable coupling. 

The laser field at frequency $\omega_L$ is formed from the superposition of three traveling plane waves of equal intensity and wave vectors ${\bf k}_i$ in the $xy$ plane. The set-up discussed in Ref.~\onlinecite{nigelnew} has triangular symmetry, with an angle of $2 \pi /3$ between the beams. The wave vectors are then ${\bf k}_1= -k/2 (\sqrt 3, 1, 0)$, ${\bf k}_2= k/2 (\sqrt 3, -1, 0)$ and ${\bf k}_3= k (0, 1, 0)$. Up to a scale factor and rotation, this is the same geometry as the asymmetric hexagonal lattice in Sec. \ref{sec:hex}. 

Each beam is linearly polarized at an angle $\theta$ to the $z$ axis, giving 
\begin{equation}
\boldsymbol{\kappa} = \kappa \sum_{i=1}^3 e^{i{\bf k}_i \cdot {\bf r}} [\cos \theta \hat{{\bf z}} +\sin \theta ( \hat{{\bf z}} \times \hat{{\bf k}_i})] ,
\end{equation}  
where $\kappa$ is the Rabi frequency of a single beam. The relative strength of the laser fields at frequencies $\omega_L$ and $\omega_L+\delta$ will henceforth be denoted as $\epsilon=E/\kappa$. 

To define the unit cell, we consider a unitary transformation: $\hat{U} \equiv$ exp$(-i{\bf k}_3 \cdot {\bf r} \hat{\sigma}_z/2)$. This gauge transformation is state dependent and leads to the Hamiltonian in Eq. (\ref{eq:couple}), where $\hat{V}' = \hat{U}^\dagger \hat{V} \hat{U}$ has the same form as $\hat{V}$, with $\kappa_0$ replaced by $\kappa_0'=e^{-i{\bf k}_3 \cdot {\bf r} }\kappa_0$. The coupling then only includes the momentum transfers ${\bf K}_{1,2} \equiv {\bf k}_{1,2}-{\bf k}_3$. These define the reciprocal lattice vectors of the largest possible Brillouin zone, and the smallest possible real-space unit cell. This unit cell is that of the hexagonal lattice discussed in Sec. \ref{sec:hex}.

\end{document}